\documentclass[sn-nature]{sn-jnl}% Math and Physical Sciences Reference Style

\newcommand{\aj}{Astron. J.}   % Astronomical Journal
\newcommand{\apj}{Astrophys. J.}   % Astrophysical Journal
\newcommand{\aap}{Astron. Astrophys.}   % Astronomy and Astrophysics
\newcommand{\mnras}{Mon. Not. R. Astron. Soc.}   % Monthly Notices of the RAS
\newcommand{\nastro}{Nat. Astron.} % Nature Astronomy
\newcommand{\sci}{Science} % Science
\newcommand{\sciadv}{Sci. Adv.} % Science Advances
\newcommand{\solphys}{Sol. Phys.}   % Solar Physics
\usepackage{xr}
\usepackage[]{draftfigure}

\begin{document}

\title[Large Exomoons unlikely around Kepler-1625\,b and Kepler-1708\,b]{Large Exomoons unlikely around Kepler-1625\,b and Kepler-1708\,b}

\author*[1]{\fnm{Ren{\'e}} \sur{Heller}}\email{heller@mps.mpg.de}

\author*[2,3]{\fnm{Michael} \sur{Hippke}}\email{michael@hippke.de}

\affil*[1]{\orgname{Max Planck Institute for Solar System Research}, \orgaddress{\street{Justus-von-Liebig-Weg 3}, \city{G{\"o}ttingen}, \postcode{37077}, \country{Germany}}}

\affil[2]{\orgname{Sonneberg Observatory}, \orgaddress{\street{Sternwartestra{\ss}e 32}, \city{Sonneberg}, \postcode{96515}, \country{Germany}}}

\affil[3]{\orgdiv{Visiting Scholar, Breakthrough Listen Group, Berkeley SETI Research Center, Astronomy Department}, \orgname{UC Berkeley}, \orgaddress{\street{501 Campbell Hall \#3411}, \city{Berkeley}, \postcode{94720-3411}, \state{CA}, \country{USA}}}

\abstract{There are more than 200 moons in our Solar System, but their relatively small radii make similarly sized extrasolar moons very hard to detect with current instruments. The best exomoon candidates so far are two nearly Neptune-sized bodies orbiting the Jupiter-sized transiting exoplanets Kepler-1625\,b and Kepler-1708\,b, but their existence has been contested. Here we reanalyse the Hubble and Kepler data used to identify the two exomoon candidates employing nested sampling and Bayesian inference techniques coupled with a fully automated photodynamical transit model. We find that the evidence for the Kepler-1625\,b exomoon candidate comes almost entirely from the shallowness of one transit observed with Hubble. We interpret this as a fitting artifact in which a moon transit is used to compensate for the unconstrained stellar limb darkening. We also find much lower statistical evidence for the exomoon candidate around Kepler-1708\,b than previously reported. We suggest that visual evidence of the claimed exomoon transits is corrupted by stellar activity in the Kepler light curve. Our injection-retrieval experiments of simulated transits in the original Kepler data reveal false positive rates of 10.9\,\% and 1.6\,\% for Kepler-1625\,b and Kepler-1708\,b, respectively. Moreover, genuine transit signals of large exomoons would tend to exhibit much higher Bayesian evidence than these two claims. We conclude that neither Kepler-1625\,b nor Kepler-1708\,b are likely to be orbited by a large exomoon.
}

\keywords{Extrasolar planets, Extrasolar moons, Transit method, Kepler Space Mission}

\maketitle

\section{Introduction}
\label{sec:introduction}

From the discovery of Jupiter's four principal moons in 1610 by Galileo Galilei \cite{Galilei1610}, which triggered the Copernican revolution, to the discovery of cryovolcanism on Saturn's moon Enceladus \cite{2006Sci...311.1393P} as evidence of ongoing liquid water-based chemistry in the outer Solar System, moons continue to deliver fundamental and fascinating insights into planetary science. The detection of moons around some of the thousands of extrasolar planets known today has thus been eagerly anticipated for over a decade now \cite{2006A&A...450..395S,2009MNRAS.400..398K,2013MNRAS.432.2549A}.

Although more than a dozen methods have been proposed to search for exomoons \cite{2018haex.bookE..35H}, the search for moons in stellar photometry of transiting planets is the only method that has been applied by several research teams \cite{2001ApJ...552..699B,2007A&A...476.1347P,2012ApJ...750..115K,2015ApJ...806...51H,2017A&A...603A.115L}. The most promising search technique seems to be photodynamical modeling \cite{2011MNRAS.416..689K,2022A&A...662A..37H}, which maximizes the signal-to-noise ratio (S/N) of any exomoon transit that might be present \cite{2022A&A...657A.119H}. No exomoon has been securely detected so far, and the main reason for this is probably that moons larger than Earth are rare \cite{2015ApJ...806...51H,2018AJ....155...36T}. For comparison, the largest moons in the solar system, Ganymede (around Jupiter) and Titan (around Saturn), have radii of about 40\,\% of the radius of the Earth. Exomoons of this size are below the detection limits even in the high-accuracy space-based photometry from the Kepler mission.

So far, two possible exomoon detections have been put forward, both of which had originally been claimed in stellar photometry from the Kepler space mission \cite{2010Sci...327..977B}. The first candidate corresponds to a Neptune-sized moon in a wide orbit around the Jupiter-sized planet Kepler-1625\,b \cite{2018AJ....155...36T}, which is in a 287\,d orbit around the evolved solar-type star Kepler-1625. The second exomoon claim has recently been announced by the same team. It is around the Jupiter-sized planet Kepler-1708\,b \cite{2022NatAsKipping}, which is in a 737\,d orbit around the solar-type main-sequence star Kepler-1708.

Given the importance of possible extrasolar moon discoveries for the field of extrasolar planets and planetary science in general, those proposed candidates call for an independent analysis. Photodynamical modelling of planet-moon transits is computationally very demanding due to the three-body nature of the star-planet-moon system and due to the complicated calculations involved in the overlapping areas of three circles \cite{Fewell2006}. Although some open-source computer code packages cover some combination of Keplerian orbital motion solvers and multi-body occultations \cite{2017ApJ...851...94L,2019AJ....157...64L}, they have not been adapted for studying exomoons. Another recently published algorithm \cite{2022AJ....164..111G} has been used to study a peculiar planet-planet mutual transit of Kepler-51\,b and d. 

Here we apply our new photodynamical model {\tt Pandora} \cite{2022A&A...662A..37H}, a publicly available open-source code written in the {\tt python} programming language, to investigate the exomoon claims around Kepler-1625\,b and Kepler-1708\,b. The main differences between {\tt Pandora} and {\tt LUNA}, photodynamical software that has previously been used for exomoon searches, are (1) {\tt Pandora}'s assumption of the small-body approximation of the planet whenever the resulting flux error is $<1$\,ppm, (2) the different treatment of the three circle intersections of the star, planet, and moon, (3) a different sampling of the posterior space ({\tt MultiNest} for {\tt LUNA} \cite{2018AJ....155...36T,2020AJ....159..142T}; {\tt UltraNest} for {\tt Pandora}), (4) a different conversion scheme between time stamps in the light curve and the true anomalies of the circumstellar and local planet-moon orbits and (5) an accelerated model throughput of {\tt Pandora} of about 4 to 5 orders of magnitude \cite{2022A&A...662A..37H}, while still keeping overall flux errors $<1$\,ppm.

\section{Results}
\label{sec:results}

\subsection{Kepler-1625\,b}
\label{sec:results_1625}

Using the data from the three transits observed with Kepler, we first masked one transit duration's worth of data to either side of the actual transit before detrending. We found this amount of data to correspond roughly to the planetary Hill sphere, which we omit from the detrending to avoid removal of any potential exomoon transit signature. We then explored three different approaches for detrending and fitting the Kepler data from stellar and systematic activity and combining it with Hubble data (Methods). The posterior sampling was achieved using the {\tt UltraNest} software \cite{2021JOSS....6.3001B}.

Approach 1 resulted in $2\log_e(B_{\rm mp})=15.9$, where $B_{\rm mp}$ is the Bayes factor for the planet-moon hypothesis over the planet-only hypothesis (Methods), signifying `decisive evidence' for an exomoon according to the Jeffreys scale (Supplementary Table~\ref{tab:Jeffreys}). In approach 2, the statistical evidence turned out to be about an order of magnitude lower in terms of $B_{\rm mp}$, with $2\log_e(B_{\rm mp})=11.2$. In approach 3, the Bayesian evidence for an exomoon was almost yet another order of magnitude lower with $2\log_e(B_{\rm mp})=7.3$, which signified `very strong evidence'. These results confirm the strong dependence of the statistical evidence of the exomoon-like signal on the detrending.

Figure~\ref{fig:Kepler1625}(a)--(d) shows 100 light curves for the combined fit of the Kepler and Hubble data based on approach 2 (orange lines) that were randomly chosen from the posterior distribution. We do not show any planet-only models from the corresponding posteriors since the weighting of the number of planet-moon models and the number of planet-only models is based on the likelihood of the models (Methods) and the planet-only interpretation is 265 times less probable than the planet-moon interpretation. We do, nevertheless, show the best fit of the planet-only model in Fig.~\ref{fig:Kepler1625}a--d for comparison (black solid line), which is important to our interpretation of the transit depth.

\subsubsection{Plausibility of transit solutions}
\label{sec:plausibility_K1625}

%%%% Fig. 1 %%%%
\begin{figure}[t]%
\centering
\includegraphics[width=1.0\textwidth]{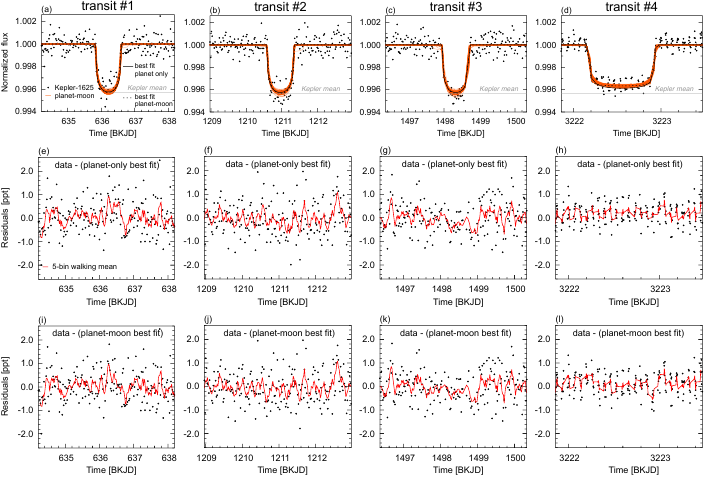}
\caption{Transit light curves of Kepler-1625\,b. Each column shows the data for one of the four transits (transits 1 to 3 from Kepler and transit 4 from Hubble), respectively. The out-of-Hill-sphere parts of the Kepler-1625\,b transit light curves were detrended using a sum of cosines, and the LDCs were used as free fitting parameters. Time in units of BKJD, which is equal to ${\rm BJD} - 2{,}454{,}833.0$\,d. (a)--(d) Orange lines visualize 100 planet-moon models that were randomly drawn from the respective posterior distributions for transit 1 (a), transit 2 (b), transit 3 (c) and transit 4 (d). Planet-only model are omitted as the corresponding Bayes factor of $2 \log_e(B_{\rm mp})=11.2$ suggests that the planet-moon interpretation is 265 times more probable than the planet-only interpretation. The best-fitting models of a planet only and of a planet with a moon are shown with solid and dashed black lines, respectively. Grey horizontal lines labeled as `Kepler mean' illustrate the mean transit depth resulting from the three transits observed with Kepler. (e)--(h) Residuals of the observed data and the best fit of the planet-only model for transit 1 (e), transit 2 (f), transit 3 (g) and transit 4 (h). Red lines denote the five-bin walking mean. (i)--(l) Residuals of the observed data and the best fit of the planet-moon model for transit 1 (i), transit 2 (j), transit 3 (k) and transit 4 (l).} 
\label{fig:Kepler1625}
\end{figure}
%%%% Fig. 1 %%%%

Although the statistical evidence is overwhelming, we noticed several things about the astrophysical plausibility of the solutions and the morphology of the transit light curves in Fig.~\ref{fig:Kepler1625}a--d that put the statistically favored planet-moon interpretation into question.

\begin{enumerate}
\item About half of the posterior models do not exhibit a single moon transit in any of the four transit epochs. This is particularly relevant since our posterior sampling with {\tt UltraNest} is very conservative in its representation of the final posteriors to assure that these posteriors are fair representations of the estimated likelihoods. The non-detection of any moon transits is not an exclusion criterion for the moon hypothesis, but it violates an important detection criterion for an exomoon interpretation \cite{2013ApJ...770..101K}. % B2 therein
\item In the other half of our posterior models that do contain moon transits, these transits occur almost exclusively in the Kepler data. This tendency for missed putative moon transits in the Hubble data has not been explicitly addressed in the literature and gives us pause to reflect on the fact that of a total of four available transits, this missed exomoon transit occurs in the one dataset that was obtained with a telescope (Hubble), unlike the remaining three transits (from Kepler).
\item From these posterior cases with a moon transit, we find only a handful of light curves with notable out-of-planetary-transit signal from the moon (Fig.~\ref{fig:Kepler1625}a--c). Instead, preferred solutions feature a moon with a small apparent deflection from the planet. This lack of solutions with moon transits at wide orbital deflections is contrary to geometrical arguments for a real exomoon. Any exomoon would spend most of its orbit in an apparently wide separation from its host planet as a result of the projection of the moon orbit onto the celestial plane \cite{2014ApJ...787...14H,2016ApJ...820...88H}. From our best fits of the orbital elements for the planet-moon models and using previously published equations for the contamination of planet-moon transits \cite{2022A&A...657A.119H}, we calculate a probability of $<10\,\%$ that such a hypothetical exomoon around Kepler-1625\,b would transit nearly synchronously with its planet during all three transits observed with Kepler. We interpret this as an artificial correction for the unconstrained stellar limb darkening, in which the ingress and egress of the moon transits are used in the fitting process to minimize the discrepancy between the data and the models.
\item The exomoon signal is almost entirely caused by the data from the Hubble observations although our model sampling of the posteriors prefers solutions in which the moon does not actually transit the star in the Hubble data. We do not find any evidence of a putative exomoon signal at 3{,}223.3\,d (BKJD) in the Hubble data (Fig.~\ref{fig:Kepler1625}d) as originally claimed \cite{2018SciA....4.1784T}. Our finding is, thus, in agreement with another study \cite{2019ApJ...877L..15K}, though these authors analyzed solely the Hubble data and not the Kepler data in a common framework.
\item The transit observed with Hubble is much shallower than the three transits observed with Kepler (Fig.~\ref{fig:Kepler1625}a--d). Our bootstrapping experiment (Methods) yields a probability of $2 \times 10^{-5}$ that the fourth transit from Hubble would have the observed transit depth, assuming the same astrophysical conditions and similar noise properties. The discrepancy can be explained as either an extrasolar moon that transits in all three transits observed with Kepler but misses the star in the single transit observed with Hubble or a wavelength dependency of the stellar limb darkening due to the different wavelength bands covered by the Kepler and Hubble instruments. Assuming only a planet and no moon as well as our best-fit estimates for the planet-to-star radius ratio, transit impact parameter and limb darkening coefficients (LDCs) for Kepler and Hubble, then we predict a transit depth of 0.99573 for the Kepler data and of 0.99634 for the Hubble data (Methods). These values are in good agreement with the observed transit depth discrepancy and offer a natural explanation that does not require a moon.
\item We confirm the previously reported transit timing variation (TTV) of the planet. Our best planet-only fit for the transit mid-point of Kepler-1625\,b at $3{,}222.55568\,(\pm0.0038)$\,d is consistent with the published value of $3{,}222.5547\,(\pm 0.0014)$\,d \cite{2019A&A...624A..95H} with a deviation of much less than the standard deviation ($\sigma$). The TTV has a discrepancy of about $3\sigma$ of the predicted transit mid-time at $3{,}222.6059\,(\pm 0.0182)$\,d using the three transits from Kepler alone. It is unclear if this timing offset was caused by a moon, by an additional, yet otherwise undetected planet around Kepler-1625 \cite{2018SciA....4.1784T,2019A&A...624A..95H,2020A&A...635A..59T} or by an unknown systematic effect. Curiously, even if we artificially correct for this TTV, the exomoon solution is still preferred over the planet-only solution with similar evidence and similar posteriors. This suggests that not the TTV but the transit depth discrepancy between the Kepler and the Hubble data is the key driver of the statistical evidence for an exomoon around Kepler-1625\,b. In other words, although the TTV between the Kepler and the Hubble data is statistically at the three-sigma level and even though the exomoon interpretation around Kepler-1625\,b hinges fundamentally on the Hubble data, the TTV effect is not as important. It is the transit depth discrepancy that causes the spurious moon signal.
\item The residual sum of squares in the combined Kepler and Hubble datasets, on a timescale of a few days, is 301.5\,ppm$^2$ for the planet-only best fit (Fig.~\ref{fig:Kepler1625}e-h) and 295.2\,ppm$^2$ for the best-fitting planet-moon model (Fig.~\ref{fig:Kepler1625}i-l). The root mean square (r.m.s) is 625.7\,ppm for the planet-only model and 619.1\,ppm for the planet-moon model, respectively. The difference in r.m.s between the models is very slim with only 6.6\,ppm. Possibly more important, this metric for the noise amplitude is larger than the depth of the claimed moon signal of about 500 ppm \cite{2020AJ....159..142T}.
\item Our properly phase-folded exomoon transit light curve has a marginal S/N of only 3.4 or 3.0, depending on the detrending. There is also no visual evidence for an exomoon transit in this phase-folded light curve of Kepler-1625\,b (Methods).
\end{enumerate}

\subsubsection{Transit injection-retrieval experiment}
\label{sec:inject_retrieve_Kepler1625}

In addition to our exomoon search around Kepler-1625\,b, we performed an injection-retrieval experiment using the original out-of-transit Kepler data of the star (Methods).

We tested 128 planet-only systems with planetary properties akin to those of Kepler-1625\,b, and we tested two families of planet-moon models, each comprising 64 simulated systems. For both simulated exomoon families, we used physical planet-moon properties corresponding to our best fit from approach 2. For one exomoon family, we tested orbital alignments like those from our best fits, whereas for the other family we tested only coplanar orbits. Moons from the coplanar family would always show transits and possibly even planet-moon eclipses, thereby increasing the statistical significance. Orbital periods for all planet-moon systems ranged between 1\,d and 20\,d.

\begin{figure*}[t]%
\centering
\includegraphics[width=1.0\textwidth]{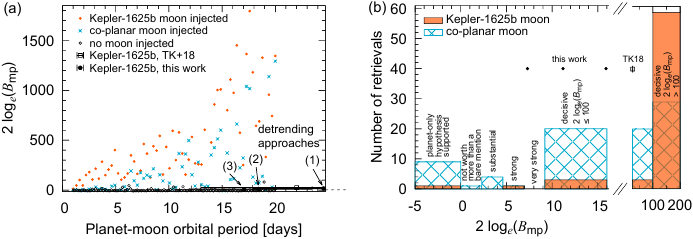}
\caption{Injection-retrieval tests of a planet-only model and two types of large moons into the out-of-transit data of the original light curve of Kepler-1625\,b. (a) Bayes factor distribution for orbital periods of the injected moons between 1\, and 20\,d. Black open circles refer to injections of planet-only models with a random spread over the planet-moon period axis. Orange points refer to injections of a Kepler-1625\,b-like planet and a moon that we parameterized according to the best-fitting posteriors of our own search. Blue dots with crosses show the outcome of simulations with a hypothetical coplanar system of a Kepler-1625\,-like planet with a large moon. The black solid circles and the black open square are the Bayes factors in this work and from TK18 \cite{2018SciA....4.1784T} (see the legend). The dashed lines in the lower right corner outside the plotting area denote the boundaries of the Jeffreys grades for $2\log_e(B_{\rm mp})$ of 0, 2.30, 4.61, 6.91, and 9.21, respectively. (b) Bayes factor histograms for the two types of injections with moons. Colors correspond to the same moon types as in (a).}
\label{fig:inj_retr_K1625}
\end{figure*}

The resulting distribution of the $2 \log_{e}(B_{\rm mp})$ values as a function of the moon's orbital period is shown in Fig.~\ref{fig:inj_retr_K1625}a. As a general observation, the Bayesian evidence increases substantially for moons in wider orbits, partly because more of the moon's in-transit data are separated from the planetary in-transit data \cite{2022A&A...657A.119H}. As an interesting side result, this is direct evidence from photodynamical modeling that a selection effect due to exomoon transit contamination by the planet will prefer exomoon discoveries in wide orbits. The Bayes factors for our own exomoon search around Kepler-1625\,b (black filled circles) and those of from previous works \cite{2018SciA....4.1784T} (empty square) are several orders of magnitude lower than those from our injection-retrieval experiments with injected moons.

%%%% Fig. 2 %%%%
%%%% Fig. 2 %%%%

Our retrievals demonstrate that our detrending does not, in the majority of all cases, erase an exomoon signal that would be present in the Kepler data. Our true positive rate, defined as `decisive' evidence on the Jeffreys scale ($2\log_e({B_{\rm mp}})>9.21$), is between 76.6\,\% and 96.9\,\%, depending on the orbital geometry of the injected planet-moon system. Details are given in Supplementary Table~\ref{tab:injretr_K1625}. For injected moons with periods near 20\,d, we find $2 \log_{e}(B_{\rm mp})$ ranging between 100 and 1{,}800. The real Kepler plus Hubble data suggests $2 \log_{e}(B_{\rm mp})$ between 7.3 (this work, detrending approach 3) and 25.9 \cite{2018SciA....4.1784T}. At the corresponding moon orbital periods of 17\,d to 24.5\,d, these $2 \log_{e}(B_{\rm mp})$ values are more compatible with our injection-retrievals of a planet-only model (black open circles). Figure~\ref{fig:inj_retr_K1625}b illustrates the same data as a $2 \log_{e}(B_{\rm mp})$ histogram, highlighting that by far most of our injected exomoons have $2 \log_{e}(B_{\rm mp})$ values larger than those found for the real transit data of Kepler-1625\,b. Importantly, in 14 out of 128 simulated planet-only transits we find $2 \log_{e}(B_{\rm mp}) > 9.21$, corresponding to a false positive rate of 10.9\,\%.

\subsection{Kepler-1708\,b}

For the two transits of Kepler-1708\,b, we tested the same three detrending and fitting approaches as for Kepler-1625\,b. Each of these approaches resulted in distinct Bayes factors when comparing the planet-only and the planet-moon models (Supplementary Table~\ref{tab:Kepler1708b}).
None of the resulting Bayes factors suggests strong evidence in favor of an exomoon interpretation. With approach 1, we obtain $2\log_e(B_{\rm mp})=-4.0$, that is to say, a $1/0.14~=~7.1$-fold statistical preference of the planet-only hypothesis. Approach 2 yields $2\log_e(B_{\rm mp})=1.0$, which is a statistical hint of an exomoon `not worth more than a bare mention' on the Jeffreys scale \cite{Jeffreys1948}. And with approach 3, we obtain $2\log_e(B_{\rm mp})=2.8$, which is substantial evidence of an exomoon around Kepler-1708\,b. Details of the posterior sampling and best-fitting model solutions are given in the Methods.

Figure~\ref{fig:Kepler1708}(a)-(b) shows a random selection of planet-only (blue) and planet-moon (orange) transit light curves from our posterior sampling with {\tt UltraNest}. This particular set of solutions was obtained with detrending approach 2. In our graphical representations, we chose to show both planet-moon solutions and planet-only solutions by weighing the number of light curves per model with the corresponding Bayes factor. In this particular case, we plot $n_{\rm p}=1/(1+B_{\rm mp})=67$\,\% of the light curves based on planet-only models and $n_{\rm mp}=1-0.5=33$\,\% with planet-moon models (Methods).

\subsubsection{Plausibility of transit solutions}
\label{sec:plausibility_K1708}

%%%% Fig. 3 %%%%
\begin{figure}[h!]%
\centering
\includegraphics[width=0.8\textwidth]{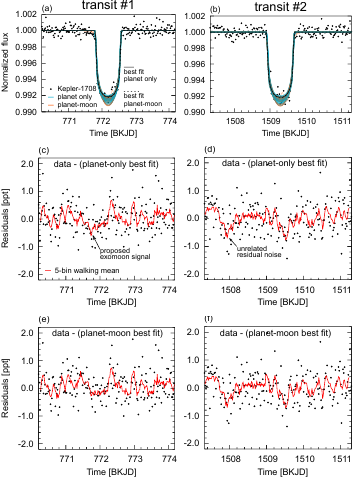}
\caption{Transit light curves of Kepler-1708\,b. The out-of-Hill-sphere parts of the Kepler-1708\,b transit light curves were detrended using a biweight filter and the LDCs were used as free fitting parameters. (a)--(b) Blue and orange lines visualize 67 planet-only models and 33 planet-moon models, respectively, that were randomly drawn from the respective posterior distributions for transit 1 (a) and transit 2 (b). The number of light curves represents the corresponding Bayes factor of $B_{\rm mp}=0.5$, which means that the planet-only interpretation is twice as probable as the planet-moon interpretation. The best-fitting models of a planet only and of a planet with a moon are shown with dashed and solid black lines, respectively. (c)--(d) Residuals of the observed data and the best fit of the planet-only model for transit 1 (c) and transit 2 (d). Red lines denote the five-bin walking mean. (e)--(f) Residuals of the observed data and the best fit of the planet-moon model for transit 1 (e) and transit 2 (f). ppt, parts per thousand.}
\label{fig:Kepler1708}
\end{figure}
%%%% Fig. 3 %%%%

We identify several aspects that are critical to the assessment of the plausibility of the exomoon hypothesis.

\begin{enumerate}\setcounter{enumi}{0}
\item It has been argued that the pre-ingress dip of transit 1 between about 771.6\,d and 771.8\,d (BKJD) cannot be caused by a star spot crossing of the planet since the planet is not in front of the star at this point \cite{2022NatAsKipping}. We second that, but we also point out that at 1{,}508\,d (BKJD), just about 1\,d before transit 2, there was a substantial decrease in the apparent stellar brightness of ${\sim}800$\,ppm (see residuals in Fig.~\ref{fig:Kepler1708}d and f) that is as deep as the suspected moon signal. This second dip near 1{,}508\,d (BKJD) also cannot possibly be related to a star spot crossing, which demonstrates that astrophysical or systematic variability may also explain the pre-ingress dip of transit 1 of Kepler-1708\,b. An exomoon is not necessary for explaining the pre-ingress variation of transit 1.
\item The residual sum of squares for the entire data in Fig.~\ref{fig:Kepler1708} is 108.4\,ppm$^2$ for the planet-only best fit and 107.7\,ppm$^2$ for the best-fitting planet-moon model. The r.m.s is 529.9\,ppm for the best-fitting planet-only model and 528.2\,ppm for the best planet-moon model. For comparison, the depth of the proposed moon transit is ${\sim}1{,}000$\,ppm and several features in the light curve have amplitudes of ${\sim}800$\,ppm on a timescale of 0.5\,d. The proposed exomoon transit signal is not distinct from other sources of variations in the light curve, which are probably of stellar or systematic origin.
\item Although we identify visually apparent dips that could be attributed to a transiting exomoon, other variations in the phase-folded light curve that cannot possibly be related to a moon cast doubt on the exomoon hypothesis (Methods).
\item Most of the claimed photometric moon signal occurs during the two transits of the planetary body, which makes it extremely challenging to discern the exomoon interpretation from limb darkening effects related to the planetary transit. This finding is reminiscent of our analysis of the transits of Kepler-1625\,b. Due to geometrical considerations it is, in fact, unlikely {\it a priori} that a moon performs its own transit in a close apparent deflection to its planet.
\item Our orbital solutions for the proposed exomoon vary substantially depending on the detrending method. As an example, the orbital period of the moon obtained from our best fits is either 12.0\,$(\pm 19.0)$\,d, 1.6\,$(\pm 5.6)$\,d, or 7.2\,$(\pm 6.2)$\,d for detrending approaches 1, 2 and 3, respectively. We verified that these are not aliases on the same orbital mean motion frequency comb but rather completely independent solutions. For a real and solid exomoon detection, we would expect that the solution is stable against various reasonable detrending methods.
\end{enumerate}

\subsubsection{Transit injection-retrieval experiment}
\label{sec:inject_retrieve_Kepler1708}

In the same manner as for Kepler-1625\,b, we performed 128 planet-only injection-retrievals and two sorts of 64 planet-moon injection-retrievals, all with orbital periods between 1\,d and 20\,d. For each injection, we used out-of-transit data of the original Kepler-1708\,b light curve from the Kepler mission.

%%%% Fig. 4 %%%%
\begin{figure*}[h]%
\centering
\includegraphics[width=1.0\textwidth]{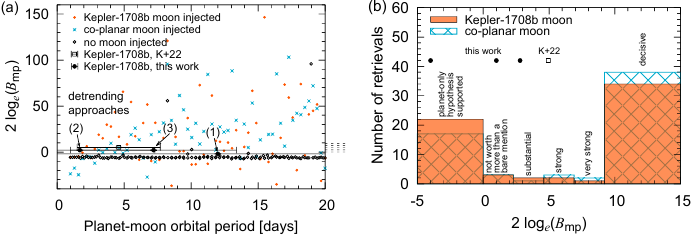}
\caption{Injection-retrieval tests of a planet-only model and two types of large moons into the out-of-transit data of the original light curve of Kepler-1708. (a) Bayes factor distribution for orbital periods of the injected moons between 1\, and 20\,d. Black open circles refer to injections of planet-only models with a random spread over the planet-moon period axis. Orange points refer to simulations with a Kepler-1708\,b-like planet and a moon that we parameterized according to the best-fit posteriors of our own search. Blue dots with crosses show the outcome of injections of a hypothetical coplanar system with a Kepler-1708\,-like planet and a large moon. The black solid and black open circles with error bars refer to the Bayes factors of this work and of K+22 (see legend). The dashed lines in the lower right corner outside the plotting area denote the boundaries of the Jeffreys grades for $2\log_e(B_{\rm mp})$ of 0, 2.30, 4.61, 6.91, and 9.21, respectively. (b) Bayes factor histograms for the two types of injections with moons. Colors correspond to the same moon types as in (a).}
\label{fig:inj_retr_K1708}
\end{figure*}
%%%% Fig. 4 %%%%

Figure~\ref{fig:inj_retr_K1708}a shows the $2 \log_e(B_{\rm mp})$ distribution resulting from our injection-retrieval tests as a function of the injected orbital period of the moon. Injected moons and real measurements for Kepler-1708\,b are color-coded as in Fig.~\ref{fig:inj_retr_K1625}. The Bayes factors that we find for the injected moons indicate `decisive' Bayesian evidence ($2 \log_e(B_{\rm mp})>9.21$) in over half of the cases and values up to ${\sim}100$ when the orbital period of the planet-moon system $P_{\rm pm} > 10$\,d. We retrieved a true positive in 34 out of 64 cases (53.1\,\%) with an injected moon like the best fit and in 38 of 64 cases (59.4\,\%) with a coplanar injected moon (Methods). Figure~\ref{fig:inj_retr_K1708}b demonstrates that both our statistical evidence and the previously found evidence \cite{2022NatAsKipping} are clearly separated from about 2/3 of the population of retrieved exomoons with injected parameters drawn from the $2\,\sigma$ intervals of our best-fitting moon model to Kepler-1708\,b. The Bayes factors of the best-fitting planet-moon and planet-only models for the real transits of Kepler-1708\,b are close to the distribution of the Bayes factors of our injected planet-only models. Our false positive rate among the planet-only injections with `decisive' evidence is $2/128 = 1.6\,\%$ (Methods and Supplementary Table~\ref{tab:injretr_K1708}).

\section{Discussion}
\label{sec:discussion}

Our unified approach for detecting exomoon transits in stellar photometry includes statistical measures, plausibility checks of the obtained solutions, visual inspection of stellar light curves and careful interpretation of the posterior samplings. This results in the following interpretation of the two exomoon candidates around Kepler-1625\,b and Kepler-1708\,b.

\subsection{Exomoon candidate around Kepler-1625\,b}

The Bayesian evidence in favor of a large exomoon around Kepler-1625\,b depends strongly on the choice of the detrending method. Although we find `very strong' to `decisive' evidence ($7.3 \lesssim 2\log_e(B_{\rm mp}) \lesssim 15.9$), some new arguments leads us to conclude that Kepler-1625\,b is not orbited by a large exomoon (Results).

Another aspect that has not been addressed explicitly before is the truncated out-of-transit baseline of the Hubble data. This has a crucial effect on the shape and the depth of the transit. The incomplete detrending necessarily leads to a mis-normalization and possibly even to the injection of false positive exomoon signals \cite{2018A&A...617A..49R}. In combination with the perils induced by the wavelength dependence of the stellar limb darkening, we think that the Hubble data of the Kepler-1625\,b transit are, therefore, not useful for an exomoon search.

In addition to the excessive statistical analysis of the light curve of Kepler-1625\,b and our inspection of the noise properties of the Kepler and Hubble light curves, there is no visual evidence of any moon transit in the data. Although this is not a decisive argument against an exomoon, since visual inspection is not an ideal tool for identifying transits nor for rejecting transits, a clear transit signal would be something that everybody would like to see for a first detection of an exomoon. In this case, the extraordinary claim of an exomoon around the giant planet Kepler-1625\,b is not supported by any visual evidence in the data of an exomoon transit.

\subsection{Exomoon candidate around Kepler-1708\,b}

The Bayesian evidence for the proposed exomoon around Kepler-1708\,b is weaker than that for Kepler-1625\,b, ranging between a support of the planet-only hypothesis and substantial evidence for an exomoon ($-4 \lesssim 2\log_e(B_{\rm mp}) \lesssim 2.8$), depending on the light curve detrending. Whichever detrending we use, we obtain consistently lower evidence for the exomoon hypothesis than the 11.9-fold preference over the planet-only hypothesis ($2\log_e(B_{\rm mp})=4.95$) as previously claimed \cite{2022NatAsKipping}. We attribute part of this disagreement to our use of the {\tt UltraNest} software when sampling the posterior space. Previous studies used {\tt MultiNest}, which may produce biased results \cite{Buchner2016} and underestimated uncertainties \cite{2020AJ....159...73N}, both of which are avoided with {\tt UltraNest} \cite{2021JOSS....6.3001B}. Beyond our Bayesian analysis, our close inspection of the transit light curve reveals several arguments that can explain the data without the need for an exomoon (Results). 

Our injection-retrieval experiments using real out-of-transit Kepler data of Kepler-1708 show that an exomoon with similar physical properties as the previously claimed exomoon would cause a much higher Bayes factor ($10 \lesssim 2\log_e(B_{\rm mp}) \lesssim 100$) than suggested by the actual data. Although this finding in itself does not mean that there is not a real exomoon in the original Kepler-1708\,b data, it makes us suspicious that of all the possible transit realizations for a given exomoon around Kepler-1708\,b, Kepler observed two transits in which the Bayesian evidence of an exomoon  is barely above the noise level. 

Finally, the false positive rate of 1.6\,\% of our injection-retrieval tests suggests that an exomoon survey in a sufficiently large sample of transiting exoplanets with similar S/N characteristics yields a large probability of at least one false positive detection, which we think is what happened with Kepler-1708\,b (Methods).

\subsection{Exomoon detection limits}

We executed additional injection-retrieval experiments to get a more general idea of exomoon detectability with current technology. Photodynamical analyses of our simulated light curves with idealized space-based exoplanet transit photometry suggests that exomoons smaller than about $0.7\,R_\oplus$ or closer than about 30\,\% Hill radii to their gas giant host planets cannot possibly be detected with Kepler-like data. For comparison, the largest natural satellite of the Solar System, Ganymede, has a radius of about $0.41\,R_\oplus$, and all of the principal moons of the Solar System gas giant planets are closer than about 3.5\,\% of their planetary host's Hill sphere.

Thus, any possible exomoon detection in the archival Kepler data or with upcoming PLATO observations will necessarily be odd when compared to the Solar System moons. In this sense, the now refuted claims of Neptune- or super-Earth-sized exomoons around Kepler-1625\,b and Kepler-1708\,b could nevertheless foreshadow the first genuine exomoon discoveries that may lay ahead.

\section{Methods}
\label{sec:methods}

\subsection{Model parameterization}

Our planet-only model has seven fitting parameters for Kepler-1708\,b and nine fitting parameters for Kepler-1625\,b. For both systems, we used the circumstellar orbital period of the planet ($P_{\rm p}$), the orbital semimajor axis ($a_{\rm p}$), the planet-to-star radius ratio ($r_{\rm p}$), the planetary transit impact parameter ($b_{\rm p}$), the time of the first planetary mid-transit ($t_{\rm 0,p}$), and two LDCs for the quadratic limb darkening law to describe the limb darkening in the Kepler band ($u_{\rm 1,K}$, $u_{\rm 2,K}$). For Kepler-1625\,b, we also require two additional LDCs to capture the limb darkening in the Hubble band ($u_{\rm 1,HST}$, $u_{\rm 2,HST}$).

It is important to note the methodological difference to the model used in the previous study that claimed the Neptune-sized exomoon around Kepler-1625\,b \cite{2018SciA....4.1784T}. That model also included a parameter to fit for any possible radius discrepancy between the Kepler and the Hubble data. Taking one step back, there are two possible reasons for a transit depth discrepancy in two different instrumental filters, for example from Kepler and Hubble. First, the planet can actually have different apparent radii in different wavelength bands, for example caused by a substantial atmosphere with wavelength-dependent opacity \cite{2002ApJ...568..377C}. Second, the wavelength dependence of stellar limb darkening can lead to different shapes and different maximum flux losses during the transit, even for a planet without an atmosphere \cite{2019A&A...623A.137H}. The first aspect of the wavelength dependence of the planetary radius was covered for Kepler-1625\,b in the first study that analyzed the combined Kepler plus Hubble data in the search for an exomoon \cite{2018SciA....4.1784T}. These authors found that the radius ratio of the planet in the Hubble and the Kepler data was ${\sim}1$, with a standard deviation of about 1\,\%. This result can be retrieved from their Table~2 (second parameter $R_{\rm p,HST}/R_{\rm p,Kep}$) and from their Fig.~S16 (parameter $p_{\rm H}/p_{\rm K}$). The largest discrepancy is found with their quadratic detrending method, which yields $R_{\rm p,HST}/R_{\rm p,Kep} = 1.009\,(+0.019, -0.017)$. The upper limit within $1\,\sigma$ is $1.009+0.019 = 1.028$. Our best fit of the planet-to-star radii ratio is $0.0581\,(\pm 0.0004)$, depending on the detrending method. To achieve a radius discrepancy of 1.028, our planet-to-star radii ratio would need to be about $0.0597 / 0.0581 \sim 1.028$ between the Kepler and the Hubble data, which is $4\,\sigma$ away from our best fit. We are, thus, sufficiently confident that we can drop the wavelength dependence of the planetary radius in our fitting procedure. As for the second aspect of the wavelength dependence of stellar limb darkening, this astrophysical phenomenon naturally reproduces the observed transit depth discrepancy plus the difference in the transit profiles, all at one go. This can be seen by comparing Fig.~\ref{fig:Kepler1625}a-c with Fig.~\ref{fig:Kepler1625}d, in which the transit in the Hubble data is fitted well with two different pairs of LDCs and without the need for a wavelength dependence of the planetary radius. All things combined, a planetary radius dependence on wavelength is not required. Instead, the wavelength dependence of stellar limb darkening can naturally explain the different transit shapes and transit depths between the Kepler and the Hubble data. This difference in our model parameterization leads to different solutions for the posteriors compared to the previous study \cite{2018SciA....4.1784T}.

Our planet-moon model includes a total of 15 fitting parameters for Kepler-1708\,b: the stellar radius ($R_{\rm s}$), two stellar LDCs to parameterize the quadratic limb darkening law ($u_{\rm 1,K}, u_{\rm 2,K}$), the circumstellar orbital period of the planet-moon barycenter ($P_{\rm b}$), the time of inferior conjunction of the first mid-transit of the planet-moon barycenter ($t_{\rm 0,b}$), the orbital semimajor axis of the planet-moon barycenter ($a_{\rm b}$), the transit impact parameter of the planet-moon barycenter ($b_{\rm b}$), the planet-to-star radii ratio ($r_{\rm p}$), the planetary mass ($M_{\rm p}$), the moon-to-star radii ratio ($r_{\rm m}$), the orbital period of the planet-moon system ($P_{\rm pm}$), the inclination of the planet-moon orbit against the circumstellar orbital plane ($i_{\rm pm}$), the longitude of the ascending node of the planet-moon orbit ($\Omega_{\rm pm}$), the orbital phase of the moon at the time of barycentric mid-transit ($\tau_{\rm pm}$) and the mass of the moon ($M_{\rm m}$). For Kepler-1625\,b we required another two LDCs for the Hubble data ($u_{\rm 1,HST}, u_{\rm 2,HST}$), making a total of 17 fitting parameters in this case. In principle, {\tt Pandora} can also model eccentric orbits, which would add another four fitting parameters (for details see ref. \cite{2022A&A...662A..37H}), but we focused on circular orbits in this study. All times are given as barycentric Kepler Julian day (BKJD), which is equal to barycentric Julian day (BJD)$-2{,}454{,}833.0$\,d. 

As our priors for the star Kepler-1625 (KIC\,4760478), we used a stellar mass of $M_{\rm s}=1.113_{-0.076}^{+0.101}\,M_\odot$ (subscript $\odot$ refers to solar values), a radius of $R_{\rm s}=1.739_{-0.161}^{+0.143}\,R_\odot$ and an effective temperature of $T_{\rm eff}=5{,}542_{-132}^{+155}$\,K, as derived from isochrone fitting \cite{2020AJ....159..280B}. %, and an age of $7.65_{-2.24}^{2.07}$\,Gyr,
For the star Kepler-1708 (KIC\,7906827), we used as our priors $M_{\rm s}=1.061_{-0.079}^{+0.073}\,M_\odot$, $R_{\rm s}=1.141_{-0.066}^{+0.073}\,R_\odot$ and $T_{\rm eff}=5{,}972_{-122}^{+126}$\,K%, age = $3.96_{-2.27}^{2.93}$\,Gyr)
 \cite{2020AJ....159..280B}.

In one of our approaches to fitting the data with {\tt Pandora}, we fixed the stellar LDCs to study the effect of stellar limb darkening on the posterior distribution and the evidence of any exomoon signal. For Kepler-1625\,b, we used two sets of LDCs. In the band of Hubble's Wide Field Camera 3, we used the same LDCs as a previous study \cite{2019ApJ...877L..15K} ($u_{\rm 1,HST}=0.216$, $u_{\rm 2,HST}=0.183$), the values of which were derived from {\tt PHOENIX} stellar atmosphere models \cite{2013A&A...553A...6H} for a main-sequence star with an effective temperature of $T_{\rm eff}=5700$\,K and with solar metallicity, [Fe/H]=0. To ensure consistency between the fixed LDCs in the Kepler and Hubble passbands, we derived the LDCs in the Kepler band from pre-computed tables \cite{2011A&A...529A..75C}, again based on {\tt PHOENIX} stellar atmosphere models for a star with $T_{\rm eff}=5700$\,K, [Fe/H]=0 and a surface gravity of $\log(g / [{\rm cm\,s}^{-2}]) = 4.5$, for which ($u_{\rm 1,K}=0.482$, $u_{\rm 2,K}=0.184$).

Although $t_{\rm 0,p}$ is the time of the first planetary mid-transit in our model parameterization, {\tt UltraNest} requires a prior ($T_0$), which we take from the literature. For Kepler-1625\,b, we use $T_0 = 636.210$\,d \cite{2018SciA....4.1784T} and for Kepler-1708\,b we use $T_0 = 772.193$\,d \cite{2022NatAsKipping} (all times in BKJD). We restricted the {\tt UltraNest} search for $t_0$ to within ${\pm}\,0.1$\,d around the prior. This yielded $t_0 = T_0 + 0.01_{-0.01}^{+0.01}$ for the planet-only model of Kepler-1625\,b and $t_0 = T_0 + 0.01_{-0.02}^{+0.02}$ for the barycenter of the planet-moon model of Kepler-1625\,b. For Kepler-1708\,b, we obtained $t_0 = T_0 - 0.01_{-0.00}^{+0.00}$ for the planet-only model and $t_0 = T_0 - 0.01_{-0.01}^{+0.01}$ for the barycenter of the planet-moon model.

The remaining planetary and orbital priors were drawn from uniform distributions.

\subsection{Light curve detrending}

Detrending has been shown to have a major effect on the statistical evidence for exomoon-like signals in transit light curves \cite{2018SciA....4.1784T}. Detrending can even inject artificial exomoon-like false positive signals in real data \cite{2018A&A...617A..49R}. Moreover, a solid case for an exomoon claim should be robust against different detrending methods. Hence, we consider the detrending part of our data analysis a crucial step and test three different approaches.

In all three detrending approaches, our {\tt Pandora} model included two stellar LDCs for the Kepler data and an independent set of two LDCs for the Hubble data, both sets of which were used to parameterize the quadratic stellar limb-darkening law.

In detrending approach 1, we fixed the four LDCs based on stellar atmosphere model calculations \cite{2011A&A...529A..75C}. The detrending of the Kepler data was done using a sum of cosines as implemented in the W{\={o}}tan software \cite{2019AJ....158..143H}, which is a re-implementation of the {\tt CoFiAM} algorithm \cite{2013ApJ...770..101K} that has previously been used to detect exomoon-like transit signals around Kepler-1625\,b and Kepler-1708\,b.

In approach 2, we explored the effect of treating the LDCs as either fixed or as free fitting parameters. We also used a sum of cosines for detrending as in approach 1, but the two sets of two LDCs were treated as free parameters during the fitting process.

In approach 3, we also used the four LDCs as free parameters but used the biweight filter implemented in {\tt W{\={o}}tan}. The biweight filter has become quite a popular algorithm for detrending stellar light curves in search for exoplanet transits since it has the highest recovery rates for transits injected into simulated noisy data \cite{2019AJ....158..143H}. Hence, we consider Tukey's biweight algorithm also a natural choice for detrending when searching exomoon transits.

Of course, more detrending methods could be explored, for example polynomial fitting \cite{2018A&A...617A..49R} and linear, quadratic or exponential fitting \cite{2018SciA....4.1784T}. As demonstrated for detrending light curves when searching for exoplanet transits \cite{2019AJ....158..143H}, an optimal detrending function that works best in every particular case may not exist for exomoons either. Hence, we restrict our study to three detrending approaches that we found to perform exquisitely in our injection-retrieval experiments, as they have low false positive and false negative rates as well as high true positive and true negative rates.

Supplementary Fig.~\ref{fig:Kepler1625_cornerplots} (for Kepler-1625\,b) shows the resulting posterior sampling from {\tt UltraNest} for detrending approach 2, as it produces the highest Bayes factor in favor of an exomoon signature. Moreover, in Supplementary Fig.~\ref{fig:Kepler1708_cornerplots} (for Kepler-1708\,b), we illustrate the {\tt UltraNest} posteriors after detrending with approach 3 for the same reason. The posterior samplings for the other two approaches appear qualitatively similar, although the exact values differ. We decided to present the maximum likelihood values and their respective standard deviations for each parameter in the column titles of these corner plots. These maximum likelihood values are different from the values that we list in Supplementary Table~\ref{tab:Kepler1625b} (for Kepler-1625\,b) and Supplementary Table~\ref{tab:Kepler1708b} (for Kepler-1708\,b), which present the mean values and standard deviations of the posterior samplings. We opted for these two different representations of the results between the corner plots and tables to give different perspectives on the non-Gaussian and often multimodal posterior samplings.

\subsection{Bayesian evidence from nested sampling}

We use the Bayes factor as our principal statistical measure to compare the planet-only and planet-moon models. The Bayes factor is defined as the ratio of marginalized likelihoods of two different models. The marginal likelihood can be viewed as the integral over the posterior density $\int d{\theta} L(D{\mid}\theta) \pi(\theta)$, where $L(D{\mid}\theta)$ is the likelihood function and $\pi(\theta)$ is the prior probability density. We define the marginal likelihood of the transit model including a moon as $Z_{\rm m}$ and the marginal likelihood of the planet-only transit model as $Z_{\rm p}$. In our work, the natural logarithm of the Bayesian evidence $\log_e(Z)$ is computed numerically for both models (and given the respective data) using {\tt UltraNest} \cite{2021JOSS....6.3001B}. Then the corresponding Bayes factor is

\begin{align}\label{eq:Bayes}
B_{\rm mp} = \frac{Z_{\rm m}}{Z_{\rm p}} = \frac{ \exp\{ \log_e(Z_{\rm m}) \} }{ \exp\{ \log_e(Z_{\rm p}) \} } = \exp\{\log_e(Z_{\rm m}) - \log_e(Z_{\rm p})\} \ ,
\end{align}

\noindent
where the $\log_e$ function refers to the natural logarithm, that is, the logarithm to base $e$ (Euler's number). In the context of previous exomoon searches, the Bayes factor ($B$) has often been quoted on a logarithmic scale as $\log_e(B)$ \cite{2018AJ....155...36T} or $2 \log_e(B)$ \cite{2018SciA....4.1784T}. On this scale a preference of the planet-only (planet-moon) model is indicated by negative (positive) values.

The Jeffreys scale \cite{Jeffreys1948} has become widely used as a tool in astrophysics to translate Bayes factors into spoken language. It has also been used in a modified form \cite{KassRaftery1995} for previous estimates of the evidence of exomoons around Kepler-1625\,b \cite{2018SciA....4.1784T} and Kepler-1708\,b \cite{2022NatAsKipping}. Although the Jeffreys scale originally referred to the evidence against the null hypothesis ($Z_0$), we adopt the equivalent perspective of the evidence in favor of the alternative hypothesis ($Z_1$), in our case the evidence for an exomoon. Hence, we use the inverse numerical values for the Bayesian factor as discussed in the appendix of Jeffreys' work \cite{Jeffreys1948}. In our terminology, $B_{10}=Z_1/Z_0$ is the Bayes factor designating the evidence in favor of $Z_1$ over $Z_0$. Our adaption of the Jeffreys scale is shown in Supplementary Table~\ref{tab:Jeffreys}, which also presents the corresponding values of $2 \log_e(B_{10})$ as well as the odds ratio in favor of the alternative hypothesis ($Z_1$).

In representing the light curves that are randomly drawn from the posterior samples of {\tt UltraNest}, we plot both planet-moon and planet-only solutions by taking into account the corresponding Bayes factor. We require that the ratio between the number of light curves with a moon ($n_{\rm mp}$) and the number of light curves based on a planet-only model ($n_{\rm p}$) is equal to the ratio of the corresponding marginalized likelihoods, $n_{\rm mp}/n_{\rm p}~=~B_{\rm mp}$. Moreover, the sum of the ratios must be $n_{\rm mp} + n_{\rm p}~=~1$. Substitution of $n_{\rm mp}$ yields $n_{\rm p}B_{\rm mp} + n_{\rm p}~=~1$, which is equivalent to $n_{\rm p}=1/(1+B_{\rm mp})$.

We utilize this conversion between the Bayes factor and the odds ratio of the evidences under investigation in Eq.~\eqref{eq:Bayes} and contextualize it as a means to assess the deviation of a particular $B$ measurement from the normal distribution of $B$ measurements, assuming that the noise is normally distributed. This evaluation is done using the error function ${\rm erf}(x)=2/\sqrt{\pi}\int_0^x dt \ e^{-t^2}$, which we compute numerically using {\tt erf()}, which is a built-in {\tt python} function in the {\tt scipy} library. Given a deviation of $n$ times the standard deviation ($\sigma$) from the mean value of a normal distribution, the value of ${\rm erf}(n/\sqrt{2})$ gives the fraction of the area under the normalized Gaussian curve that is within the error bars, in particular for $n=1$ one obtains the well-known ${\rm erf}(1/\sqrt{2})=66.8$\,\%.

The odds can then be calculated as $O = 1/(1- {\rm erf}(n/\sqrt{2}) )$, and with Eq.~\eqref{eq:Bayes} we have $\log_e(B)=\log_e(O)$. Then a $3\sigma$ detection is signified by $\log_e(B)~{\geq}~5.91$, a $4\sigma$ detection by $\log_e(B)~{\geq}~9.67$ and a $5\sigma$ detection by $\log_e(B)~{\geq}~14.37$ (Supplementary Fig.~\ref{fig:logB_sigma}). These numbers are in agreement with the results from previous 200 injection-and-retrieval tests \cite{2022NatAsKipping}. From their sample of planet-only injections into the out-of-transit Kepler light curve of Kepler-1708\,b, these authors found one false positive exomoon detection with $\log_e(B)>5.91$. For comparison, we found that the odds for such a $3\sigma$ detection are 1/370, and so for 200 retrievals with an injected planet-only model, we would expect $200/370=0.54$ false positives, which is 1 when rounded to the next full integer.

% python> import scipy
% python> from scipy import *
% python> from numpy import log, sqrt
% python> O = 1/(1-scipy.special.erf(3 / sqrt(2))); print("%.2f" % (log(O)))
% 5.91

\subsection{Convergence of nested sampling}

For nested sampling, we used {\tt UltraNest} with multimodal ellipsoidal region and region slice sampling. The Mahalanobis measure is used to define the distance between start and end points of our walkers. The strategy terminates as soon as the measure exceeds the mean distance between pairs of live points. Specifically, {\tt UltraNest} integrates until the live point weights are insignificant ($<0.01$). In different experiments, we used static and dynamic sampling strategies with 800 to 4{,}000 active walkers and always required 4{,}000 points in each island of the posterior distribution before a sample was considered independent. All experiments yielded virtually identical results, showing excellent robustness. In addition, we performed 1{,}000 injection-retrieval experiments to ensure that the recovery pipeline was robust.

Likelihood surface exploration is sufficiently complete after about $10^8$ model evaluations for our data (Supplementary Fig.~\ref{fig:convergence}), whereas approximately $10^9$ model evaluations yielded only marginal gains. Many other sampling strategies, such as reactive nested sampling or the use of correlated model parameters, led to slower convergence by up to three orders of magnitude. Moreover, the {\tt MultiNest} software previously used for planet-only and planet-moon model evaluations of the transit light curves of Kepler-1625\,b and Kepler-1708\,b has been shown to yield biased results \cite{Buchner2016} and to systematically underestimate uncertainties in the best fit parameters \cite{2020AJ....159...73N}. These two key problems of {\tt MultiNest} are avoided in {\tt UltraNest} \cite{2021JOSS....6.3001B}. Our corresponding {\tt UltraNest} sampling of the models generated with {\tt Pandora} took 14\,hrs on a single core of an AMD Ryzen 5950X processor.

With regards to our {\tt UltraNest} fits of Kepler-1625\,b, detrending approach 1 resulted in more than $2.5 \times 10^8$ planet-moon model evaluations, approach 2 in over $1.3 \times 10^9$ planet-moon model evaluations and approach 3 in almost $2.3 \times 10^8$ planet-moon model evaluations. For the {\tt UltraNest} sampling of the Kepler-1708\,b data after detrending with approaches 1, 2 and 3, we generated $1.6 \times 10^8$, $2.3 \times 10^8$ and $1.7 \times 10^8$ planet-moon model evaluations, respectively.

For comparison, typical nested sampling of $5\times10^8$ model evaluations (Supplementary Fig.~\ref{fig:convergence})
takes 9 hours on a single 4.8\,GHz core of an Intel Core i7-1185G7 at a typical speed of $15{,}000$ model evaluations per second.

\subsection{Exomoon detectability}

In view of now several exomoon candidate claims near the detection limit, the general question about exomoon detectability in space-based stellar photometry arises. Due to the high computational demands of exoplanet-exomoon fitting \cite{2011MNRAS.416..689K,2022A&A...662A..37H}, this question cannot be addresses in an all-embracing manner for all possible transit surveys, cadences, system parameters, etc. Nevertheless, we executed a limited and idealized injection-retrieval experiment to determine the smallest possible moons that are detectable in Kepler-like data of (hypothetical) photometrically quiescent stars.

All stars exhibit intrinsic photometric variability, which is caused by magnetically-induced star spots, p-mode oscillations, granulation and other astrophysical processes. Moreover, any observation -- even high-accuracy space-based photometry -- comes with instrumental noise components from the readout of the charged coupled devices (CCDs), long-term telescope drift, short-term jitter, intra-pixel non-uniformity, charge diffusion, loss of the CCD quantum efficiency, etc. After modeling and removing the instrumental effects, the photometrically most quiet stars with a Kepler magnitude ${\it Kp}<12.5$ from the Kepler mission have been shown to exhibit a combined differential photometric precision (CDPP) over 6.5\,hr of about 20\,ppm \cite{2015AJ....150..133G}. Given that the nominal long cadence of the Kepler mission is 29.4\,min and that the S/N scales with the square root of the number of data points, this corresponds to an amplitude of 72\,ppm per data point, although great care should be taken when interpreting the CDPP as a measure of stellar activity \cite{2015AJ....150..133G}.

In our pursuit to identify the idealized scenarios in which exomoons can be found, that is to say, to identify the smallest exomoons possible, we consider a nominal Neptune-sized planet in a 60\,d orbit around a Sun-like star, corresponding to a semimajor axis of 0.3\,AU. To some extend, we have in mind the most abundant population of warm mini-Neptune exoplanets that this hypothetical planet could represent. Over 2, 3, and 4 years, such a planet would show 12, 18, and 24 transits, respectively. We also envision an exomoon around this planet, for which we test different physical radii and orbital periods around the planet. In the following, we find it helpful to refer to the extent of the moon orbit in units of the Hill radius ($R_{\rm Hill}=a_{\rm b}(M_{\rm p}/[3M_{\rm s}])^{1/3}$), which can be considered as a sphere of the gravitational dominance of the planet. Moons in a prograde orbital motion, which orbit the planet in the same sense of rotation as the direction of the planetary spin, become gravitationally unbound beyond $\sim0.4895\,R_{\rm Hill}$ \cite{2006MNRAS.373.1227D}. Retrograde moons, for comparison, can be gravitationally bound even with semimajor axes up to ${\sim}0.9309\,R_{\rm Hill}$ \cite{2006MNRAS.373.1227D}, depending on the orbital eccentricity. For comparison, the Galilean moons reside within 0.8\,\% and 3.5\,\% of Jupiter's Hill radius, Titan sits at 1.8\,\% of Saturn's Hill radius, and Triton orbits at 0.3\,\% of Neptune's Hill radius. The Earth's Moon has an orbital semimajor axis of about $0.26\,R_{\rm Hill}$.

In our experiment, we test exomoon injections throughout the entire Hill radius, which corresponds to an orbital period of about 33\,d. For all our simulations, we used the {\tt Pandora} software \cite{2022A&A...662A..37H} to generate planet-moon transit models at 30\,min cadence to which we added normally distributed white noise as described. For each test case, we simulated a total of 18 transits over a nominal mission duration of 3\,yr, representative of a Kepler-like space mission. The upcoming PLATO mission, for example, will observe two long-observation phase fields for either 2\,yr + 2\,yr or for 3\,yr + 1\,yr, respectively, in the hunt for Earth-like planets around Sun-like stars \cite{2014ExA....38..249R,2022A&A...665A..11H}. We then used the {\tt UltraNest} software to populate the posteriors in the parameter space of both the planet-only and the planet-moon models and computed the Bayes factors, as in the main part of this study for Kepler-1625\,b and Kepler-1708\,b. The whole exercise was then repeated for moon orbital periods between 1\,d and 33\,d and moon radii between $0.5\,R_\oplus$ and $1.0\,R_\oplus$. We define an exomoon recovery as an {\tt UltraNest} detection of the injected signal with $2\log_{\rm e}(B_{\rm mp})>9.21$, corresponding to decisive evidence on the Jeffreys scale.

Supplementary Fig.~\ref{fig:injection}a shows one simulated transit of our hypothetical warm Neptune-sized exoplanet and its Earth-sized moon around a Sun-like star in the white noise limit as described. The moon transit is barely visible by the human eye and is statistically insignificant. After 18 transits, however, the transit becomes statistically significant and is even detectable in the phase-folded light curve of the planet-moon barycenter as the orbital sampling effect \cite{2014ApJ...787...14H,2016ApJ...820...88H}, see Supplementary Fig.~\ref{fig:injection}b. Supplementary Fig.~\ref{fig:recovery} shows the distribution of our recoveries in the parameter plane spanned by the moon radius and the moon's orbital semimajor axis in units of $R_{\rm Hill}$. As a main result, we find that moons smaller than about $0.7\,R_\oplus$ are barely detectable even for these idealized cases with completely inactive stars and a total of 18 transits for a given planet-moon system. Moreover, the recovery rate drops to zero for orbits closer than about $0.3\,R_{\rm Hill}$, which corresponds to orbital periods $<5.5$\,d. This latter finding is in line with recent findings for the preservation of the exomoon in-transit signal being favored in wide exomoon orbits \cite{2022A&A...657A.119H}.

\subsection{Injection-retrieval tests}

The purpose of our injection-retrieval experiments for the observational data of Kepler-1625\,b and Kepler-1708\,b is twofold. First, we wanted to control the ability of our detrending approach to preserve any exomoon transit signal in those cases where an exomoon is, indeed, present in the data. Second, we wanted to quantify the probability that our detrending approach induces a false exomoon signal in those cases in which no injected exomoon transit is actually present.

Our experiment began with the preparation of light curve segments that contain only stellar plus instrumental and systematic effects but no known planetary transits or possible moon transits. We removed the known planetary transits as well as two day segments before and after each planetary mid-transit time, respectively. For each injection of a planet-moon transit with {\tt Pandora}, a random time in the remaining Kepler light curve is chosen. We then extracted a segment of 5\,d around each injected mid-transit time for further use and validated that no more than five data points were missing to avoid using gaps in our experiment.

In the next step, we created synthetic models with {\tt Pandora}. These were either planet-only models or models with planet-moon systems. As for the planet-only injections, for both Kepler-1625\,b and Kepler-1708, we performed a total of 128 exomoon searches in the light curve segments that contained only a planetary transit injection, with planetary properties drawn from our planet-only solutions for Kepler-1625\,b or Kepler-1708\,b, respectively. We chose negligible moon masses and radii, and the planet-moon orbital periods were chosen successively between 1\, and 20\,d with a constant step size of $(20-1)\,{\rm d}/128=0.148375$\,d. Strictly speaking, the choice of these periods is irrelevant since no moons were effectively injected in the planet-only data, but this arrangement of the data simplified the use with {\tt Pandora} and it aided the representation of the $2\log_e(B_{\rm mp})$ distribution from the planet-only injections in Figs.~\ref{fig:inj_retr_K1625} and \ref{fig:inj_retr_K1708}.

As for the exomoon injections, we distinguished two sorts of exomoons. For each type, there were included 64 simulations on a grid of orbital periods between 1\, and 20\,d and a constant step size of $(20-1)\,{\rm d}/64=0.0297$\,d. For both Kepler-1625\,b and Kepler-1708\,b, we assumed one scenario of a moon in a coplanar orbit, that is to say, with $i_{\rm pm}=0^\circ$ and $\Omega_{\rm pm}=0^\circ$, but with randomized orbital phase offsets ($\tau_{\rm m}$). This setup ensured that the were moon transits during every planetary transit and that planet-moon eclipses occured occasionally, a scenario that should increase the statistical signal of the moon. In a second scenario, we injected a planet and moon with the same radii and orbital distance but now $i_{\rm pm}$ and $\Omega_{\rm pm}$ were drawn randomly from within the $2\,\sigma$ confidence interval of our posterior distributions obtained using detrending approach 2. This scenario is representative of the best-fitting exomoon solutions for Kepler-1625\,b and Kepler-1708\,b and helped us to assess the true positive and false negative rates of our real exomoon search in the actually observed transits.

We injected these synthetic models in independent runs. In each run, a randomly chosen Kepler data segment was multiplied by the synthetic signal. Then the stellar and instrumental noise was detrended using {\tt W{\={o}}tan}'s implementation of Tukey's biweight filter \cite{2019AJ....158..143H} with a window size of three times the planetary transit duration while masking the actual planetary transit before the calculation of the trend.

Finally, we ran {\tt UltraNest} twice for each injected transit sequence, once with a planet-moon model and once with a planet-only model. The Bayes factor was then calculated in the form $2\log_e(B_{\rm mp})$.

\subsubsection{Injection-retrieval for Kepler-1625\,b}
\label{sec:inject_K1625}

The statistics of the original exomoon claim around Kepler-1625\,b \cite{2018AJ....155...36T} was determined using the {\tt LUNA} photodynamical model code \cite{2011MNRAS.416..689K} together with {\tt MultiNest} sampling \cite{2009MNRAS.398.1601F} in a Bayesian framework. This resulted in $2\log_e(B_{\rm pm})=20.4$ and an interpretation of `strong evidence' of an exomoon according to the Kass \& Raftery scale \cite{KassRaftery1995}. During their investigations of the Hubble follow-up observations, the authors re-examined the Kepler data and noticed a substantial decrease of the Bayes factor to $2\log_e(B_{\rm pm})=1$, which means that the evidence of an exomoon was essentially gone in the Kepler data.

The reason was found in an update of the Kepler Science Processing Pipeline of the Kepler Science Operations Center (SOC) from version 9.0 (v.9.0) to v.9.3. Although the initial exomoon claim study \cite{2018AJ....155...36T} used data from SOC pipeline v.9.0, the subsequent study \cite{2018SciA....4.1784T} used Kepler data from SOC pipeline v.9.3. The previous exomoon claim has now been explained as being a mere systematic effect in the Kepler data. Ironically, when adding the new transit data from Hubble observations, a new exomoon-like signal was found with $2\log_e(B_{\rm pm}) = 11.2$ or $2\log_e(B_{\rm pm}) = 25.9$, depending on the method used for detrending the out-of-transit light curve. The claimed moon was now in a very wide orbit at $\approx$~40 planetary radii from the planet and with an orbital period of $P_{\rm pm}~=~22_{-9}^{+17}$\,d, although the posterior distribution of $P_{\rm pm}$ was highly multimodal \cite{2018SciA....4.1784T}.

Previous studies \cite{2020AJ....159..142T} also describe a transit depth of 500\,ppm for an exomoon candidate around Kepler-1625\,b in the Hubble data. Their authors argued that if this feature were due to star spots rather than due an exomoon, the depth of the signal should be about 650\,ppm in the Kepler data, given the different bandpass response functions of Kepler and Hubble. They fitted box-like transit models to 100{,}000 out-of-transit regions of the Kepler data of Kepler-1625b and found that 3.8\,\% of the experiments resulted in box-like transits deeper than 650\,ppm (depth $>650$\,ppm) and that 3.5\,\% of the tests produced negative (inverted) transits with amplitudes below 650\,ppm (depth $<650$\,ppm).

Their injection-recovery tests of simulated data with only white noise resulted in similar though slightly smaller rates of such false positives with a similar symmetrical behavior of positive and negative transits. The authors of these previous studies concluded that the spurious detections in the real and simulated Kepler data are, thus, due to Gaussian (white) noise rather than to time-correlated noise from star spots or other periodic stellar activity.

Our own injection-retrieval experiments for Kepler-1625\,b were not restricted to the assumption of white noise. Instead, we used transit-free light curve segments from the original Kepler data of Kepler-1625 as described above. We used the fourth transit from Hubble as is, as there was not enough out-of-transit Hubble data to inject and retrieve artificial transits and to do proper detrending for recovery.

Figure~\ref{fig:inj_retr_K1625} shows the results of our injection-retrieval tests for Kepler-1625\,b. Of the 128 injections of planet-only models (black circles), 96 scatter between $2\log_e(B_{\rm mp})=-0.13$ and $-7.49$. With 114 systems showing a Bayes factor lower than our `decisive' detection limit of $2\log_e(B_{\rm mp})=9.21$, we determine a true negative rate of $89.1\,\%$ and a false positive rate of 10.9\,\%.

Of our 64 simulated planet-moon systems that were parameterized according to our {\tt UltraNest} posteriors (orange dots), 61 (95.3\,\%) showed $2 \log_{e}(B_{\rm mp}) > 15.9$. More generally, we retrieved $62/64=96.9\,\%$ of all moons with $2\log_e(B_{\rm mp})>9.21$, 59 of which even had $2\log_e(B_{\rm mp})>100$.

From the injected transit models that included a moon on a coplanar orbit (pale blue dots with crosses), 45 (70.3\,\%) had $2 \log_{e}(B_{\rm mp}) > 15.9$, as obtained with our detrending approach 1 of the original Kepler data. We also measure a true positive rate ($2\log_e(B_{\rm mp})>9.21$) of $49/64=76.6\,\%$, of which 29 successful retrievals signified $2\log_e(B_{\rm mp})>100$.

\subsubsection{Injection-retrieval for Kepler-1708\,b}
\label{sec:inject_K1708}

The exomoon claim paper for Kepler-1708\,b proposes a super-Earth-sized moon with a radius of $R_{\rm m}=2.61_{-0.43}^{+0.42}\,R_\oplus$ at a distance of $11.7_{-2.2}^{+3.9}\,R_{\rm p}$ and with an orbital period of $P_{\rm pm}=4.6_{-1.8}^{+3.1}$\,d. The authors of that paper calculated a Bayes factor of $B_{\rm mp}=11.9$, which means $2 \log_e(B_{\rm mp}) = 4.95$ \cite{2022NatAsKipping} and `strong evidence'. The authors performed 200 injections of a planet-only signal, in which they found 40 systems with $2 \log_e(B_{\rm mp}) > 0$ and two systems with $2 \log_e(B_{\rm mp}) > 4.61$ (their Fig.~3, but note the abscissa scaling and the limit at $\log_e(B_{\rm mp}) > 2.3$). 

Figure~\ref{fig:inj_retr_K1708} presents the outcome of all these simulations. Black open circles represent the 128 planetary transit injections without a moon, the $2\log_e(B_{\rm mp})$ values of which are scattered between about -1.5 and -7.4. Orange points represent the exomoon-exoplanet injections that we sampled from the $2\,\sigma$ confidence interval of our best fit using detrending approach 2. Blue points with crosses refer to the coplanar exomoon-exoplanet injections. For comparison, we plotted the measurements for the proposed exomoon signal around Kepler-1708\,b from previous work \cite{2022NatAsKipping} and from this work (Supplementary Table~\ref{tab:Kepler1708b}). In 22 of the 64 tests (34.4\,\%) with an injected moon that was parameterized from the $2\,\sigma$ posteriors, we found $2 \log_e(B_{\rm mp}) < 0$, that is, the moon signal was completely lost. In 39 out of 64 cases (60.9\,\%), we found a $2 \log(B_{\rm mp})$ value that is higher than the value of 2.8 that we derived by fitted the LDCs using a biweight filter. In 17 out of 64 cases (26.6\,\%) with coplanar planet-moon orbits, we found $2 \log_e(B_{\rm mp}) < 0$ and the moon signal was completely lost. In 44 out of 64 cases (68.8\,\%), we recovered the injected moon that was parameterized akin to the candidate around Kepler-1708\,b with a $2 \log_e(B_{\rm mp})$ value larger than the value of 2.8 that we obtained by fitting LDCs and using a biweight filter for detrending.

In summary, the actual value of $2 \log_e(B_{\rm mp}) = 2.8$ for the proposed exomoon candidate is rather small compared to the values that we typically obtain from our injection-retrieval tests. Whenever there is really a moon in the data, it can be found with higher confidence than the proposed candidate in most cases. The Bayes factor of the candidate in the real Kepler data is also suspiciously close to the distribution of systems for which there was actually no moon present (Fig.~\ref{fig:inj_retr_K1708}).

In two of our 128 cases that included only planetary transits, we obtained $2 \log_e(B_{\rm mp}) > 9.21$. That is, our false negative rate was 1.6\,\%. This value is compatible with the false positive rate of $1.0_{-1.0}^{+0.7}\,\%$ reported by \cite{2022NatAsKipping}. This finding highlights an interesting aspect that goes beyond the detection of an exomoon claim around Kepler-1708\,b. Our false positive rate is equivalent to a probability of $(1-2/128)^1=98.4\,\%$ that we do not detect a false positive exomoon in a Kepler-1708\,b-like transit light curve. In two exomoon searches, the probability that we would not produce a single false positive would be $(1-2/128)^2=96.9\,\%$. After $n$ searches, the probability of not detecting a false positive would be $(1-2/128)^n$, and after 70 attempts the probability of having no false positive is 33.2\,\%. In turn, the probability of having at least one false positive after 70 exomoon searches is $1 - (1-2/128)^{70}=66.8\,\%$. Of course, this estimate is only applicable to stellar light curves with comparable stellar activity and noise characteristics. However, we find this an interesting side note given that the exomoon claim paper of Kepler-1708\,b included a sample of 70 transiting planets \cite{2022NatAsKipping}. From this perspective, the detection of a false positive giant exomoon around Kepler-1708\,b is, maybe, not as surprising.

\subsection{Phase-folded transit light curves}

We artificially re-added the planetary contribution to the combined planet-moon transit, which is not just a simple addition of a single planetary transit model, due to the possible planet-moon eclipses, but requires careful modeling with our photodynamical exoplanet-exomoon transit simulator {\tt Pandora} \cite{2022A&A...662A..37H}. Supplementary Fig.~\ref{fig:Kepler1625_fold} illustrates that there is no appealing visual evidence of an exomoon transit in the observations of Kepler-1625. The depth of the putative exomoon transit varies substantially between 500\,ppm for approach 2 and 100\,ppm for approach 3, but the S/N was also marginal at $<3.4$ or $<3.0$ for all four transits, depending on the detrending approach. 

In both Supplementary Figs.~\ref{fig:Kepler1708_fold}a (detrending approach 2) and \ref{fig:Kepler1708_fold}b (detrending approach 3), we see the folding of the two proposed exomoon transits around zero mid-transit time. However, we also see another dip of almost similar depth at about $-1.5$\,d before the planetary mid-transit of transit 2 (orange dots), which corresponds to the dip at 1{,}508\,d (BKJD) mentioned above in our discussion of Supplementary Fig.~\ref{fig:Kepler1708}. So, for Kepler-1708\,b there actually is a visual hint of a stellar flux decrease in addition to the transit of the planet. However, its proximity in the light curve to another substantial variation in the light curve casts a serious doubt on the exomoon nature of the stellar flux decrease.

Hence, neither in the phase-folded light curve of the barycenter of Kepler-1625\,b and its proposed moon nor for that of Kepler-1708\,b did we identify any visually apparent variation that could be exclusively explained by an exomoon transit.

\subsection{Transit depth discrepancy of Kepler-1625\,b}

To assess the probability that the observed discrepancy for the transit depths of Kepler-1625\,b in the Kepler and Hubble data could be due to a statistical variation, we executed a bootstrapping experiment. We simulated the three transits observed with Kepler based our measurements of mid-transit flux of $0.99571$, $0.99566$, and $0.99567$, respectively, and with formal uncertainties of 0.0001. These mid-transit fluxes and the uncertainties were chosen as mean values and standard deviations from which we drew 10 million randomized samples for each of the three transits.

The resulting histogram is shown in Supplementary Fig.~\ref{fig:bootstrapping}. The transit depth of transit 4 from Hubble is indicated with an arrow at 0.99610 with a formal uncertainty of roughly 30\,ppm. From the total of 30 million realizations, we measured a fraction of $2 \times 10^{-5}$ with a transit depth greater than or equal to the observed transit depth from Hubble. It is, thus, highly unlikely that the observed transit depth discrepancy in the Kepler versus the Hubble data is a statistical variation, assuming normally distributed errors. Instead, an astrophysical origin, red noise, or an unknown cause are required as an explanation.

We advocate for an astrophysical explanation that is well known in stellar physics and that does not require an exomoon. The radial profile of the apparent stellar brightness (or stellar intensity), known as stellar limb-darkening profile, depends on the wavelength band that a star is observed in. This effect was originally observed for the Sun \cite{1977SoPh...51...25P}. Limb-darkening profiles can be described well by {\it ad hoc} limb-darkening laws, for which we use a quadratic limb-darkening law that is parameterized by two LDCs. When the stellar transit of an extrasolar planet is observed in two different filters, then the resulting LDCs and transit depth can vary substantially \cite{2019A&A...623A.137H}, whereas the transit impact parameter and the planet-to-star radii ratio must, of course, be the same.

Assuming circular orbits, the mid-transit depth ($\delta$) can be expressed in terms of the minimum in-transit flux ($f_{\rm min}$) as $\delta = 1 - f_{\rm min}$, so that we can predict the minimum in-transit flux with $f_{\rm min} = 1 -\delta$ if we can predict $\delta$. Using the expression of the transit depth as a function of the transit overshoot factor from the light curve ($o_{\rm LC}$) \cite{2019A&A...623A.137H} (Eq.~(1) therein), we have

\begin{equation}
\delta = (1 + o_{\rm LC}) {\Big (} \frac{R_{\rm p}}{R_{\rm s}} {\Big )}^2 \ .
\end{equation}

\noindent
Using Eq.~(3) in ref.~\cite{2019A&A...623A.137H} and our best-fitting estimates from the planet-only model with $(R_{\rm p}/R_{\rm s})=0.05818$, an impact parameter $b_{\rm p}=0.11$, and LDCs for Kepler ($u_{\rm 1,K}=0.42$, $u_{\rm 2,K}=0.41$) and Hubble ($u_{\rm 1,H}=0.12$, $u_{\rm 2,H}=0.21$), we predict a transit depth of 0.99573 for the Kepler data and of 0.99634 for the Hubble data. These values are in good agreement with the transit depth discrepancy that we actually observe (Fig.~\ref{fig:Kepler1625}). The transit depth discrepancy between the Kepler and the Hubble data can, thus, be readily explained by the wavelength dependence of stellar limb darkening, and it does not require an exomoon.

\subsection{Methodological comparison to previous studies of Kepler-1625\,b}

Although there has not been any follow-up study to test the exomoon claim around Kepler-1708\,b, various papers have analyzed the Kepler and Hubble transit data of Kepler-1625\,b. Here we provide a brief historical summary of the debate around Kepler-1625\,b and its proposed exomoon candidate and give an overview of the methodological differences between our study and previous studies.

The initial statistical `decisive evidence' of an exomoon with $2\log_e(B_{\rm pm})=20.4$ ($B_{\rm pm}$) was based on three transits available in archival Kepler data from 2010 to 2013 \cite{2018AJ....155...36T}. In a subsequent study \cite{2018SciA....4.1784T}, the authors noticed that the evidence of an exomoon in the Kepler data was gone ($2\log_e(B_{\rm pm})=1$), which they attributed to an update of the Kepler Science Processing Pipeline of the Kepler Science Operations Center (SOC) from version 9.0 (v.9.0) to v.9.3. The original exomoon claim around Kepler-1625\,b has thus been explained as a systematic effect. A new exomoon claim was made by the same authors based on new observations of a fourth transit observed with the Hubble Space Telescope from 2017 \cite{2018SciA....4.1784T}, % (Program GO 15149: PI: A. Teachey).
with $2\log_e(B_{\rm pm})$ ranging between 11.2 and 25.9 for various detrending methods used for the light curve segments. Curiously, the Hubble observations showed a TTV compared to the strictly periodic transits from Kepler, which could in principle be caused by the gravitational pull of a giant moon on the planet. % around the local planet-moon barycenter.
Reported TTVs range between 77.8\,min \cite{2018SciA....4.1784T} and $73.728\,(\pm 2.016)$\,min \cite{2019A&A...624A..95H}. The strong dependence of the statistical evidence on the details of the data preparation has, however, questioned the exomoon interpretation around Kepler-1625\,b \cite{2018A&A...617A..49R,2019A&A...624A..95H,2019ApJ...877L..15K}.

\begin{enumerate}
    \item Our study applies the same software and the same kind of injection-retrieval test to the transits of both Kepler-1625\,b and Kepler-1708\,b in a unified framework.
    \item Refs. \cite{2018A&A...617A..49R} and \cite{2019A&A...624A..95H} used a numerical scheme that was hardcoded specifically to the case of exoplanet-exomoon transit simulations for Kepler-1625\,b. Their code is not public, and thus, it has been challenging for the community to reproduce their results.
    \item Refs. \cite{2018AJ....155...36T} and \cite{2018A&A...617A..49R} studied only the three transits from the Kepler mission because the follow-up transit observations with Hubble were not available at the time. In our study, we combine data of four transits from the Kepler and Hubble missions.
    \item Ref. \cite{2019ApJ...877L..15K} studied only the single transit observed with Hubble but none of the three transits from the Kepler mission. 
    \item Refs. \cite{2018AJ....155...36T} and \cite{2018A&A...617A..49R} used Kepler data from the Kepler SOC pipeline v.9.0. As first noted by ref. \cite{2018SciA....4.1784T}, the previously claimed exomoon signal around Kepler-1625\,b that was present in the Simple Aperture Photometry measurements in the discovery paper \cite{2018AJ....155...36T} vanished after the upgrade of Kepler's SOC pipeline to v.9.3. We use data from Kepler's SOC pipeline v.9.3 in our new study. These new data has also been used by refs. \cite{2018SciA....4.1784T}, \cite{2019A&A...624A..95H}, and \cite{2020AJ....159..142T}.
    \item Refs. \cite{2018A&A...617A..49R}, \cite{2019A&A...624A..95H}, and \cite{2019ApJ...877L..15K} used the differential Bayesian information criterion for the planet-only and the planet-moon models, whereas refs. \cite{2018AJ....155...36T}, \cite{2018SciA....4.1784T}, and \cite{2020AJ....159..142T} used the Bayes factor. We also use the Bayes factor in our study.
    \item Refs. \cite{2018A&A...617A..49R} and \cite{2019A&A...624A..95H} used Markov chain Monte Carlo sampling of the posterior distribution, which is prone to becoming trapped in local regions of the parameter space. Refs. \cite{2018SciA....4.1784T} and \cite{2020AJ....159..142T} used the {\tt MultiNest} software for the posterior sampling, which can introduce biases in the fitting process \cite{Buchner2016} and which underestimates the resulting best fit uncertainties \cite{2020AJ....159...73N}. In contrast to all those previous studies, we used the {\tt UltraNest} software for posterior sampling, which avoids these problems \cite{2021JOSS....6.3001B}.
    \item Only one previous study of the transit light curve of Kepler-1625\,b featured injection-retrieval experiments \cite{2020AJ....159..142T}. The methods of the injection-retrieval experiment used in this previous study assumed box-like transits and were, thus, less realistic than those we applied. Moreover, we disagree with the conclusions of these authors about the occurrence rate of false positive exomoon-like transit signals in the Kepler data (Sect.~\ref{sec:inject_K1625}).
\end{enumerate}

\subsection{Transit animations}
\label{sec:animations}

For both Kepler-1625\,b and Kepler-1708\,b, we generated video animations of the best-fitting planet-moon solutions in the posterior distributions. These animations were generated with the {\tt Pandora} software using the model parameterization for the maximum likelihood provided by our {\tt UltraNest} sampling. At the times of the transit midpoints of the respective planet-moon barycenter, we exported a screenshot, the results of which are shown in Supplementary Fig.~\ref{fig:animation}. The colors of the stars Kepler-1625 and Kepler-1708 were chosen automatically in {\tt Pandora} to reflect the stellar colors as they would be perceived by the human eye, according to previously published digital color codes of main-sequence stars \cite{2021AN....342..578H}. We increased the frame rate to five times its default value, which is one frame every 30\,min or 48 frames per day. Our animations, thus, come with 240 frames of simulated data per day and they are played at a rate of 60 frames per second.

As shown by the corresponding corner plots in Supplementary Figs.~\ref{fig:Kepler1625_cornerplots} and \ref{fig:Kepler1708_cornerplots}, the posterior distributions are very scattered and any moon solutions are ambiguous at best. As we have discussed in the main text, it is much more likely that there is no large exomoon around either planet. So the purpose of these animations is mostly a general illustration of planet-moon orbital dynamics during transits as well as an interpretation of the transit light curves (and potentially debugging) rather than to represent the actual transit events. If {\tt Pandora}'s animation functionality were to be used to visualize actual transit events, then the posterior distributions would need to be much more well-confined and the Bayes factors of the solutions would need to be much higher (and, thus, the solution more convincing) than for Kepler-1625\,b or Kepler-1708\,b.

\backmatter

\bmhead{Data availability}
The Kepler light curves are available for download at the Mikulski Archive for Space Telescopes (MAST) at \href{https://archive.stsci.edu/kepler/publiclightcurves.html}{https://archive.stsci.edu/kepler/publiclightcurves.html}. The raw data of the Hubble observations of the October 2017 transit of Kepler-1625\,b are available at \href{https://www.stsci.edu/hst/observing/program-information}{https://www.stsci.edu/hst/observing/program-information} under program ID GO 15149 (PI A.~Teachey).

\bmhead{Code availability}
{\tt Pandora} is open-source and available from \href{https://github.com/hippke/Pandora}{https://github.com/hippke/Pandora}. {\tt W{\={o}}tan} is open-source and available from \href{https://github.com/hippke/wotan}{https://github.com/hippke/wotan}. {\tt UltraNest} (Copyright 2014-2020, Johannes Buchner) is open-source and available from \href{https://johannesbuchner.github.io/UltraNest}{https://johannesbuchner.github.io/UltraNest}.

\bmhead{Acknowledgments}
The authors are thankful to Sascha Grziwa and the other, anonymous, reviewer for their constructive reviews of the manuscript. They also thank  G.~Bruno for providing them with the extracted Hubble light curve of the October 2017 transit of Kepler-1625\,b. R.~H. acknowledges support from the German Aerospace Agency (Deutsches Zentrum f\"ur Luft- und Raumfahrt) under PLATO Data Center grant 50OO1501.

\bmhead{Author contributions}
R.~H. devised the research question, performed the analytical calculations and wrote the manuscript. M.~H. executed the light curve detrending, the model fitting and the nested sampling. Both authors contributed equally to the arrangements of the figures and tables and to interpreting the results.

\bmhead{Competing interests}
The authors declare that they have no competing interests.

\hfill \break \noindent
Correspondence and requests for materials should be addressed to R.~H. (\href{heller@mps.mpg.de}{mailto:heller@mps.mpg.de}) or M.~H. (\href{michael@hippke.org}{mailto:michael@hippke.org}).

\clearpage

\bibliographystyle{sn-nature}
%\bibliography{bibliography}

% ==== copy-pasted content of the bbl file from bibtex ====>

% <=== copy-pasted content of the bbl file from bibtex ====

\renewcommand{\figurename}{Supplementary Fig.}
\renewcommand{\tablename}{Supplementary Table}
\renewcommand\thefigure{\arabic{figure}}
\renewcommand\thetable{\arabic{table}}
\setcounter{figure}{0}
\setcounter{table}{0}

\clearpage

\clearpage

%%%% Supplementary Table 1 %%%%
\begin{table}[h]
\begin{center}
\caption{The Jeffreys scale of Bayesian evidence in favor of the alternative hypothesis.}
\label{tab:Jeffreys}%
\setlength{\tabcolsep}{3pt}
\begin{tabular}{@{}l|l|l|l@{}}
\toprule
Bayes factor range\footnotemark[1] & $2 \log_e(B_{10})$  & Odds & Evidence\footnotemark[1] in favor of $Z_1$\\
\midrule
\hspace{-0.02cm} $<10^{0/2}$                               & $<0$  & \hspace{0.18cm} $<1:1$                    & null hypothesis $Z_0$ supported  \\
\hspace{0.31cm} $10^{0/2}$ -- $10^{1/2}$ & \hspace{0.273cm} $0$ -- $2.30$ & \hspace{0.50cm} $1:1$ -- $3.16:1$      & not worth more than a bare mention  \\
\hspace{0.31cm} $10^{1/2}$ -- $10^{2/2}$ & $2.30$ -- $4.61$ &  \hspace{0.13cm} $3.16:1$ -- $10:1$                & substantial  \\
\hspace{0.31cm} $10^{2/2}$ -- $10^{3/2}$ & $4.61$ -- $6.91$ & \hspace{0.36cm} $10:1$ -- $31.6:1$  & strong  \\
\hspace{0.3cm} $10^{3/2}$ -- $10^{4/2}$  & $6.91$ -- $9.21$  & \hspace{0.13cm} $31.6:1$ -- $100:1$                & very strong  \\
\hspace{-0.01cm} $>10^{4/2}$                              & $>9.21$  & $>100:1$                                    & decisive \\
\botrule
\end{tabular}
%\footnotetext{(any additional notes)}
\footnotetext[1]{Bayes factor ($B_{10}$) ranges and wording of evidence after Jeffreys [31]. %\cite{Jeffreys1948}.
From top to bottom, Jeffreys referred to these ranges as Grades 0, 1, 2, 3, 4, and 5.}
\end{center}
\end{table}
%%%% Supplementary Table 1 %%%%

\clearpage

%%%% Supplementary Fig. 1 %%%%
\begin{figure*}[h!]%
\centering
\includegraphics[width=1\textwidth]{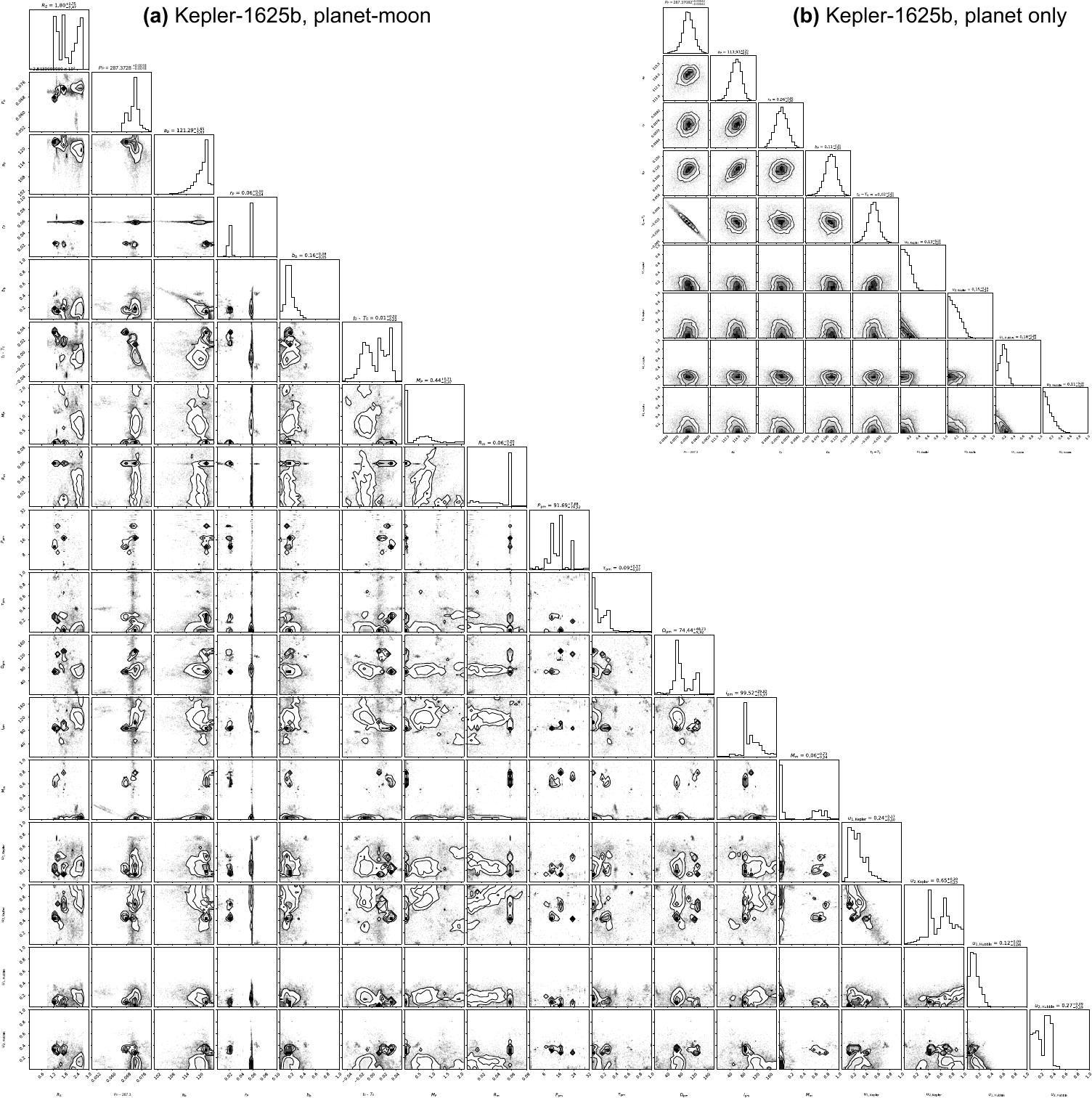}
\caption{Posterior distribution for Kepler-1625\,b for the combined data from Kepler (three transits) and Hubble (one transit). Out-of-transit detrending was performed using a sum of cosines, the light curve fitting was performed with {\tt Pandora} using the stellar LDCs as free parameters, and the sampling with {\tt UltraNest}. Planet and moon masses are given in fractions of Jupiter's mass. The stellar radius is given in units of the solar radius. Planet and moon radii are given in stellar radii. Orbital periods are given in units of days. Column titles denote the maximum likelihood values and their standard deviation. (a) Planet-moon model. (b) Planet-only model.}
\label{fig:Kepler1625_cornerplots}
\end{figure*}
%%%% Supplementary Fig. 1 %%%%

\clearpage

%%%% Supplementary Table 2 %%%%
\begin{sidewaystable}[h!]
\sidewaystablefn%
\tiny
\begin{center}
\begin{minipage}{\textheight}
\caption{Mean and standard deviation derived from the posterior distribution of our {\tt Pandora} fitting of the transit data of Kepler-1625\,b with {\tt UltraNest}.}
\label{tab:Kepler1625b}
\setlength{\tabcolsep}{3pt}
\begin{tabular}{lcc|cc|cc|cc|cc|cc}
\toprule
          & \multicolumn{2}{c}{Planet, LD fix\footnotemark[1], cos\footnotemark[2]} & \multicolumn{2}{c}{Planet+moon, LD fix, cos} & \multicolumn{2}{c}{Planet, LD free, cos} & \multicolumn{2}{c}{Planet+moon, LD free, cos} & \multicolumn{2}{c}{Planet, LD free, Biw\footnotemark[3]} & \multicolumn{2}{c}{Planet+moon, LD free, Biw} \\\cmidrule{2-3} \cmidrule{4-5} \cmidrule{6-7} \cmidrule{8-9} \cmidrule{10-11} \cmidrule{12-13}
Parameter &      Mean & Std. dev.     &      Mean & Std. dev.   &    Mean & Std. dev.     &     Mean\footnotemark[4] & Std. dev.     &     Mean & Std. dev.     &     Mean & Std. dev. \\
\midrule
$P_{\rm p/b} ({\rm d})$\footnotemark[5]  & 287.37082 & 0.00042 & 287.3728 & 0.0018 & 287.37095 & 0.00044 & 287.3722 & 0.0019 & 287.37062 & 0.00042 & 287.3712 & 0.0019 \\
$a_{\rm b} (R_{\rm s})$ & 122.6 & 1.5 & 122.4 & 1.3 & 122.43 & 1.81 &  121.04 & 3.75 & 122.1 & 2.1 & 121.2 & 2.9 \\
$r_{\rm p}$ & 0.05817 & 0.00031 & 0.050 & 0.015 & 0.05818 & 0.00049 & 0.029 & 0.018  & 0.05804 & 0.00049 & 0.047 & 0.027 \\
$b_{\rm b}$ & 0.102 & 0.076 & 0.15 & 0.16 & 0.11 & 0.08 & 0.14 & 0.11 & 0.128 & 0.093 & 0.146 & 0.095 \\
$t_{\rm 0,b}\,{\rm (d)}-T_0$ & 0.0154 & 0.0038 & 0.0206 & 0.0058 & 0.0141 & 0.0039 & 0.014 & 0.0091 & 0.0132 & 0.0037 & 0.0049 & 0.0091 \\
$u_{\rm 1,K}$ &  &  &  &  & 0.42 & 0.15 & 0.25 & 0.17 & 0.37 & 0.15 & 0.35 & 0.16 \\
$u_{\rm 2,K}$ &  &  &  &  & 0.41 & 0.25 & 0.56 & 0.26 & 0.46 & 0.25 & 0.47 & 0.26 \\
$u_{\rm 1,HST}$ &  &  &  &  & 0.13 & 0.08 & 0.17 & 0.09 & 0.13 & 0.08 & 0.15 & 0.07 \\
$u_{\rm 2,HST}$ &  &  &  &  & 0.21 & 0.14 & 0.15 & 0.12 & 0.21 & 0.14 & 0.13 & 0.08 \\
$R_{\rm s} (R_\odot)$ &  &  & 1.59 & 0.38 &  &  & 1.714 & 0.38 & & & 1.62 & 0.66 \\
$M_{\rm p} (M_{\rm J})$ &  &  & 1.37 & 0.67 &  &  & 0.39 & 0.57 & & & 0.58 & 0.73\\
$r_{\rm m}$ &  &  & 0.030 & 0.016 &  &  & 0.048 & 0.018 & & & 0.039 & 0.025 \\
$P_{\rm m} ({\rm d})$ &  & & 24.5 & 6.7 &  &  & 18.4 & 6.3 & & & 17.1 & 6.4 \\
$\tau_{\rm m}$ &  & & 0.33 & 0.22 &  &  & 0.60 & 0.33 & & & 0.63 & 0.27\\
$\Omega_{\rm m} ({\rm deg})$ &  &  & 111 & 17 &  &  & 89 & 33 & & & 60 & 29 \\
$i_{\rm m} ({\rm deg})$ &  & & 89.5 & 5.9 &  &  & 95 & 20 & & & 110 & 39 \\
$M_{\rm m} (M_{\rm J})$ & &  & 0.58 & 0.63 &  &  & 1.15 & 0.61 & & & 0.63 & 0.47 \\
\midrule
Evaluations & 403,051 &  & 254,385,520 &  & 1,034,355 &  & 1,343,246,640 &  & 1,103,578 &  & 229,599,810 & \\
$\log_e(Z)$ & -569.972 & 0.178 & -562.006 & 0.367 & -571.095 & 0.208 & -565.516 & 0.183 & -613.344 & 0.208 & -609.680 & 0.230 \\
$2 \log_e(B_{\rm mp})$ &  &  & 15.9 & &  &  & 11.2 & & &  & 7.3 & \\
\botrule
\end{tabular}
\footnotetext[1]{LDCs for the quadratic limb darkening law fixed to ($u_{\rm 1,K}=0.482$, $u_{\rm 2,K}=0.184$) in the Kepler passband and to ($u_{\rm 1,HST}=0.216$, $u_{\rm 2,HST}=0.183$) in the Hubble passband (see Methods).}
\footnotetext[2]{Detrending with sum of cosines.}
\footnotetext[3]{Detrending with biweight filter.}
\footnotetext[4]{In this best fit solution the moon is more massive than the planet. This is an artefact of the fitting method, in which the roles of the planet and the moon are symmetric. This solution is still physically plausible.}
\footnotetext[5]{For planet-only models this parameter is $P_{\rm p}$, for planet-moon models it is $P_{\rm b}$.}
\footnotetext{As the prior for our {\tt UltraNest} search for $t_0$ we used $T_0 = 636.210$\,d [27].}% \cite{2018SciA....4.1784T}.}
\footnotetext{In all the fits, the three transits from the Kepler mission (SOC pipeline v9.3) were combined with the single transit observed with Hubble. The two pairs of LDCs for the Kepler and Hubble data were fitted independently.}
\end{minipage}
\end{center}
\end{sidewaystable}
%%%% Supplementary Table 2 %%%%

\clearpage

%%%% Supplementary Fig. 2 %%%%
\begin{figure*}[h]%
\centering
\includegraphics[width=1\textwidth]{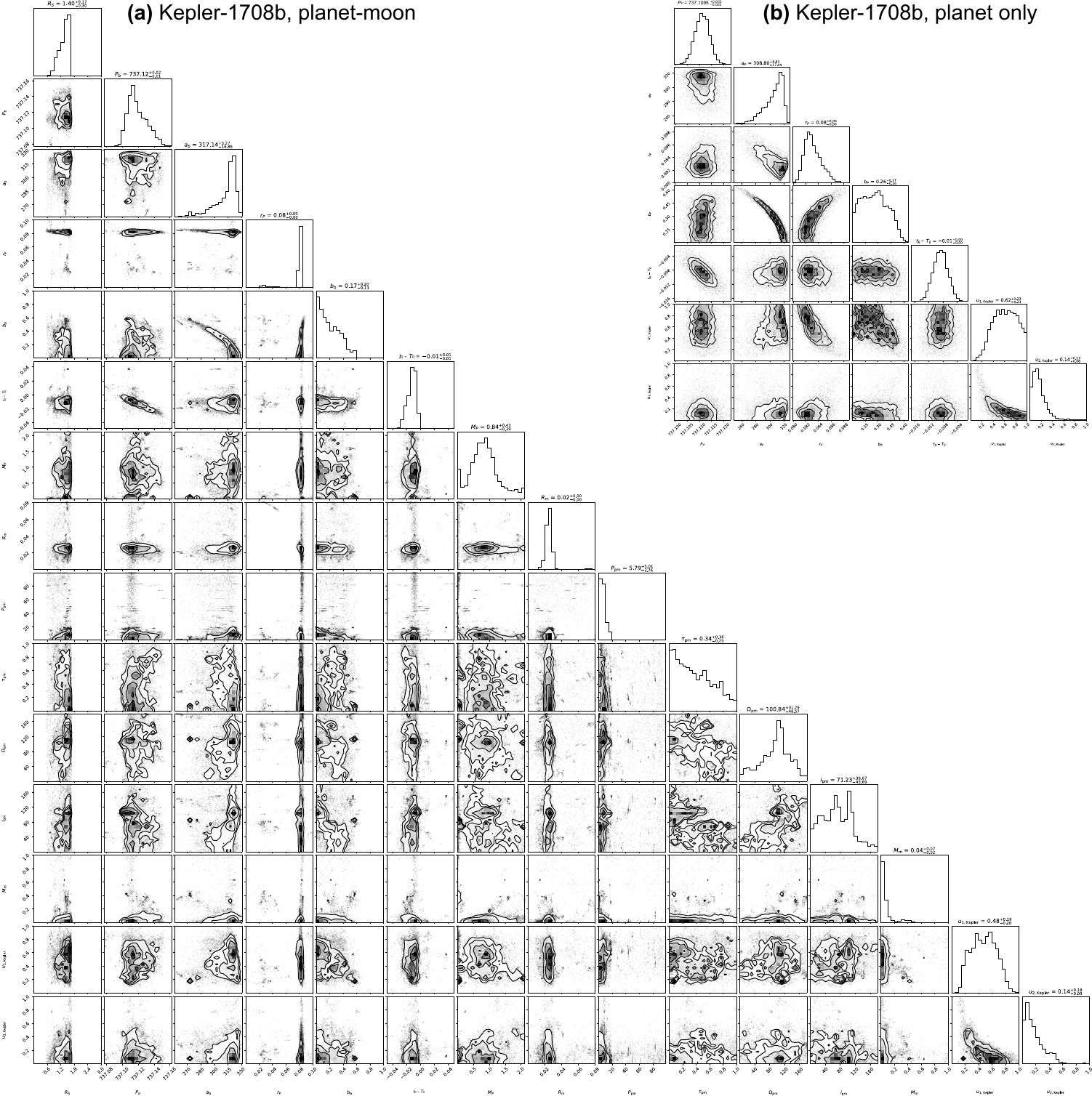}
\caption{Posterior distribution for Kepler-1708\,b using two transits available from Kepler. The out-of-transit detrending was performed using a biweight filter, the light curve fitting was performed with {\tt Pandora}, and the sampling with {\tt UltraNest}. Stellar LDCs were fitted rather than fixed. Planet and moon masses are given in fractions of Jupiter's mass. The stellar radius is given in units of the solar radius. Planet and moon radii are given in stellar radii. Orbital periods are given in units of days. Column titles denote the maximum likelihood values and their standard deviation. (a) Planet-moon model. (b) Planet-only model.}
\label{fig:Kepler1708_cornerplots}
\end{figure*}
%%%% Supplementary Fig. 2 %%%%

\clearpage

%%%% Supplementary Table 3 %%%%
\begin{sidewaystable}[h]
\sidewaystablefn%
\tiny
\begin{center}
\begin{minipage}{\textheight}
\caption{Mean and standard deviation derived from the posterior distribution of our {\tt Pandora} fitting of the transit data of Kepler-1708\,b with {\tt UltraNest}.}
\label{tab:Kepler1708b}
\setlength{\tabcolsep}{3pt}
%\begin{tabular*}{\textwidth}{lcc|cc|cc|cc}
\begin{tabular}{lcc|cc|cc|cc|cc|cc}
\toprule
          & \multicolumn{2}{c}{Planet, LD fix\footnotemark[1], Cos\footnotemark[2]} & \multicolumn{2}{c}{Planet+moon, LD fix, cos} & \multicolumn{2}{c}{Planet, LD free, cos} & \multicolumn{2}{c}{Planet+moon, LD free, cos} & \multicolumn{2}{c}{Planet, LD free, Biw\footnotemark[3]} & \multicolumn{2}{c}{Planet+moon, LD free, Biw} \\\cmidrule{2-3} \cmidrule{4-5} \cmidrule{6-7} \cmidrule{8-9}  \cmidrule{10-11} \cmidrule{12-13}
Parameter &      Mean & Std. dev.     &      Mean\footnotemark[4] & Std. dev.   &    Mean & Std. dev.     &     Mean\footnotemark[4] & Std. dev. &     Mean & Std. dev.     &     Mean & Std. dev. \\
\midrule
$P_{\rm p/b} ({\rm d})$\footnotemark[5] & 737.1095 & 0.003 & 737.128 & 0.02 & 737.1096 & 0.003 & 737.154 & 0.015 & 737.1111 & 0.003 & 737.119 & 0.012 \\
$a_{\rm b} (R_{\rm s})$ & 311.6 & 9.9 & 318.8 & 7.8 & 307.0 & 12.0 & 308.8 & 7.6 & 310.0 & 11.0 & 312 & 12.0 \\
$r_{\rm p}$ & 0.08292 & 7.4E-4 & 0.034 & 0.031 & 0.083 & 0.0013 & 0.072 & 0.017 & 0.08305 & 8.3E-4 & 0.078 & 0.013 \\
$b_{\rm b}$ & 0.22 & 0.12 & 0.17 & 0.11 & 0.24 & 0.14 & 0.261 & 0.073 & 0.23 & 0.13 & 0.2 & 0.15 \\
$t_{\rm 0,b}\,{\rm (d)}-T_0$ & -0.0079 & 0.0021 & -0.018 & 0.012 & -0.0081 & 0.0021 & -0.0304 & 0.0089 & -0.0085 & 0.0021 & -0.0128 & 0.0065 \\
$q_1$ &  &  &  &  & 0.62 & 0.2 & 0.579 & 0.078 & 0.5 & 0.29 & 0.48 & 0.18 \\
$q_2$ &  &  &  &  & 0.16 & 0.12 & 0.26 & 0.1 & 0.51 & 0.29 & 0.18 & 0.14 \\
%$R_{\rm s}$ &  &  & 901486452 & 1.54910483E8 &  &  & 645806037 & 1.29143432E8 &  &  & 943380305 & 1.47985856E8 \\
$R_{\rm s} (R_\odot)$ &  &  & 1.30 & 0.22 & 1.29 & 0.22 & 0.93 & 0.19 &  &  & 1.36  &  0.21 \\
$M_{\rm p} (M_{\rm J})$ &  &  & 0.14 & 0.34 &  &  & 1.0606 & 0.3681 &  &  & 0.8631 & 0.4573 \\
$r_{\rm m}$ &  &  & 0.076 & 0.016 &  &  & 0.035 & 0.017 &  &  & 0.028 & 0.014 \\
$P_{\rm m} ({\rm d})$ &  &  & 12 & 19.0 &  &  & 1.6 & 5.6 &  &  & 7.2 & 6.2 \\
$\tau_{\rm m}$ &  &  & 0.36 & 0.24 &  &  & 0.64 & 0.12 &  &  & 0.38 & 0.27 \\
$\Omega_{\rm m} ({\rm deg})$ &  &  & 82 & 53.0 &  &  & 100 & 19.0 &  &  & 95 & 39.0 \\
$i_{\rm m} ({\rm deg})$ &  &  & 104 & 56.0 &  &  & 72 & 14.0 &  &  & 73 & 40.0 \\
$M_{\rm m} (M_{\rm J})$ &  &  & 0.8453 & 0.4769 &  &  & 2.1010 & 1.4035 &  &  & 0.0761 & 0.1342 \\
\midrule
Evaluations & 417{,}518 &  & 159{,}338{,}332 &  & 1{,}233{,}462 &  & 232{,}805{,}871 &  & 1{,}112{,}688 &  & 165{,}259{,}346 & \\
$\log_e(Z)$ & -249.116 & 0.163 & -251.121 & 0.367 & -249.513 & 0.178 & -249.015 & 0.226 & -262.724 & 0.212 & -261.315 & 0.337 \\
$2 \log_e(B_{\rm mp})$ &  &  & -4.0 &  &  &  & 1.0 &  &  &  & 2.8 & \\
\botrule
\end{tabular}
\footnotetext{All fits based on Kepler data from SOC pipeline v9.3.}
\footnotetext[1]{LDCs fixed to $u_1=0.4582$, $u_2=0.1974$.}
\footnotetext[2]{Detrending with sum of cosines.}
\footnotetext[3]{Detrending with biweight filter.}
\footnotetext[4]{In this best fit solution the moon is more massive than the planet. This is an artefact of the fitting method, in which the roles of the planet and the moon are symmetric. This solution is still physically plausible.}
\footnotetext[5]{For planet-only models this parameter is $P_{\rm p}$, for planet-moon models it is $P_{\rm b}$.}
\footnotetext{As the prior for our {\tt UltraNest} search for $t_0$ we used $T_0 = 772.193$\,d [17].}% \cite{2022NatAsKipping}.}
\footnotetext{In two cases throughout this table, the error bars on the planet or moon masses imply negative masses. These solutions are a mathematical artefact of the fitting procedure and not physically plausible.}
\end{minipage}
\end{center}
\end{sidewaystable}
%%%% Supplementary Table 3 %%%%

\clearpage

%%%% Supplementary Fig. 3 %%%%
\begin{figure*}[h]
\centering
\includegraphics[width=.55\linewidth]{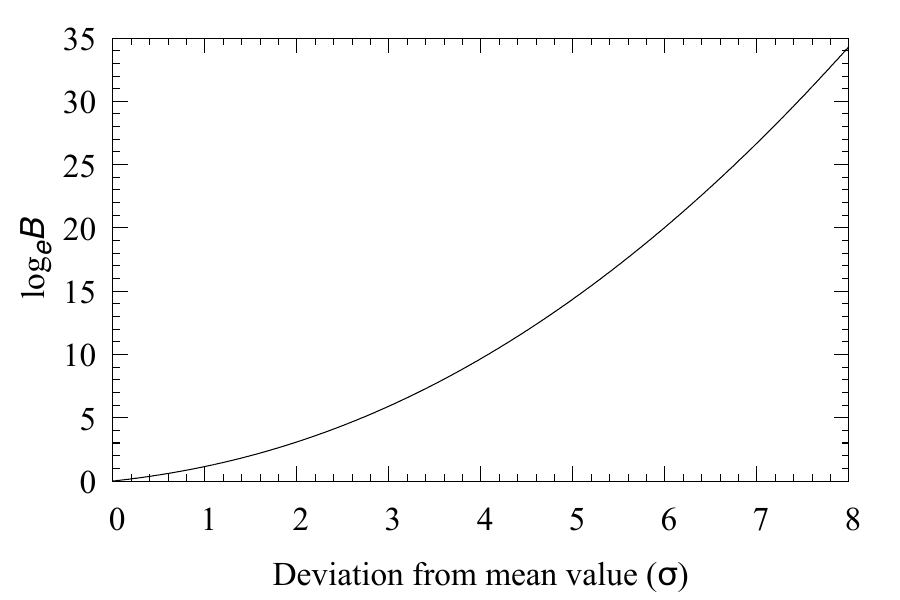}
\caption{Relation between the Bayes factor (based on the evidence of two models) and the standard deviation of a particular measurement from the mean. Normally distributed noise is assumed.}
\label{fig:logB_sigma}
\end{figure*}
%%%% Supplementary Fig. 3 %%%%

\clearpage

%%%% Supplementary Fig. 4 %%%%
\begin{figure*}[h]%
\centering
\includegraphics[width=.995\textwidth]{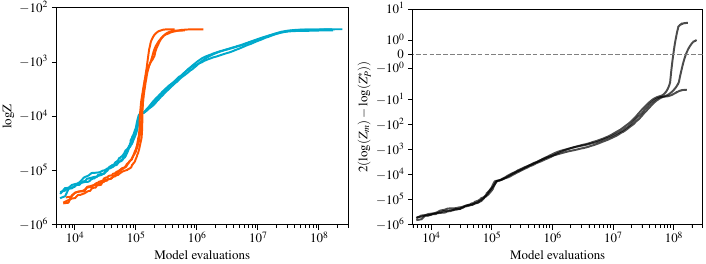}
\caption{Left: Typical convergence as a function of the number of model evaluations. In all experiments, nested sampling was performed with {\tt Pandora} and {\tt UltraNest} and identical parameters for Kepler-1708 b (i) with a slice sampler, 800 walkers, and 4000 points for independent samples. The blue curves represent planet+moon models (13 or 15 free parameters), the red planet-only models (5 to 7 free parameters). The realizations differ slightly in their choice of LDCs and detrending. Planet-with-moon models require about two orders of magnitude more model evaluations, typically a few times $10^8$. Repeating experiments with strictly identical parameters show the same dispersion due to the random nature of nested sampling. Right: Evolution of (positive or negative) evidence for the planet-with-moon model, compared to the planet-only model, as a function of model evaluations. Again, $\gtrsim 10^8$ model evaluations are required to acquire evidence for the more complex model.}
\label{fig:convergence}
\end{figure*}
%%%% Supplementary Fig. 4 %%%%

\clearpage

%%%% Supplementary Fig. 5 %%%%
\begin{figure*}[h]%
\centering
\includegraphics[width=.99\textwidth]{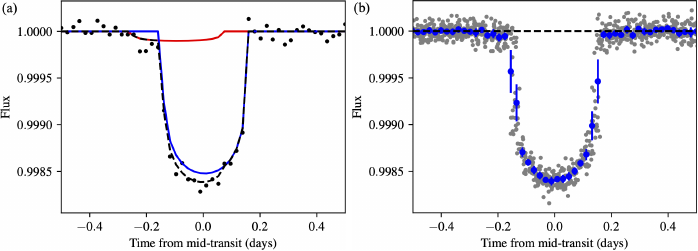}
\caption{Simulated transit of an exoplanet-exomoon system. The star is assumed to be photometrically inactive and the white noise corresponds to the minimum amplitude for Sun-like stars from the Kepler mission. The Earth-sized moon is in a 17\,d orbit around a Neptune-sized planet, both in 60\,d orbit at 0.3\,AU around a Sun-like star. (a) The contributions to the stellar dimming by the planet and the moon are shown with blue and red solid lines, respectively. The resulting transit model is shown with a black solid line. The simulated 30\,min cadence data including noise is shown with black dots. (b) Phase-folded light curve of the planet-moon barycenter after 18 transits over 3\,yr. Blue dots with error bars denote the 30\,min binned data with standard deviations. Gray dots are the original, phase-folded data points, and the dashed horizontal line shows the 100\,\% flux level of the star. The pre-ingress and post-egress flux decrease is due to the orbital sampling effect.}
\label{fig:injection}
\end{figure*}
%%%% Supplementary Fig. 5 %%%%

\clearpage

%%%% Supplementary Fig. 6 %%%%
\begin{figure*}[h]%
\centering
\includegraphics[width=.64\textwidth]{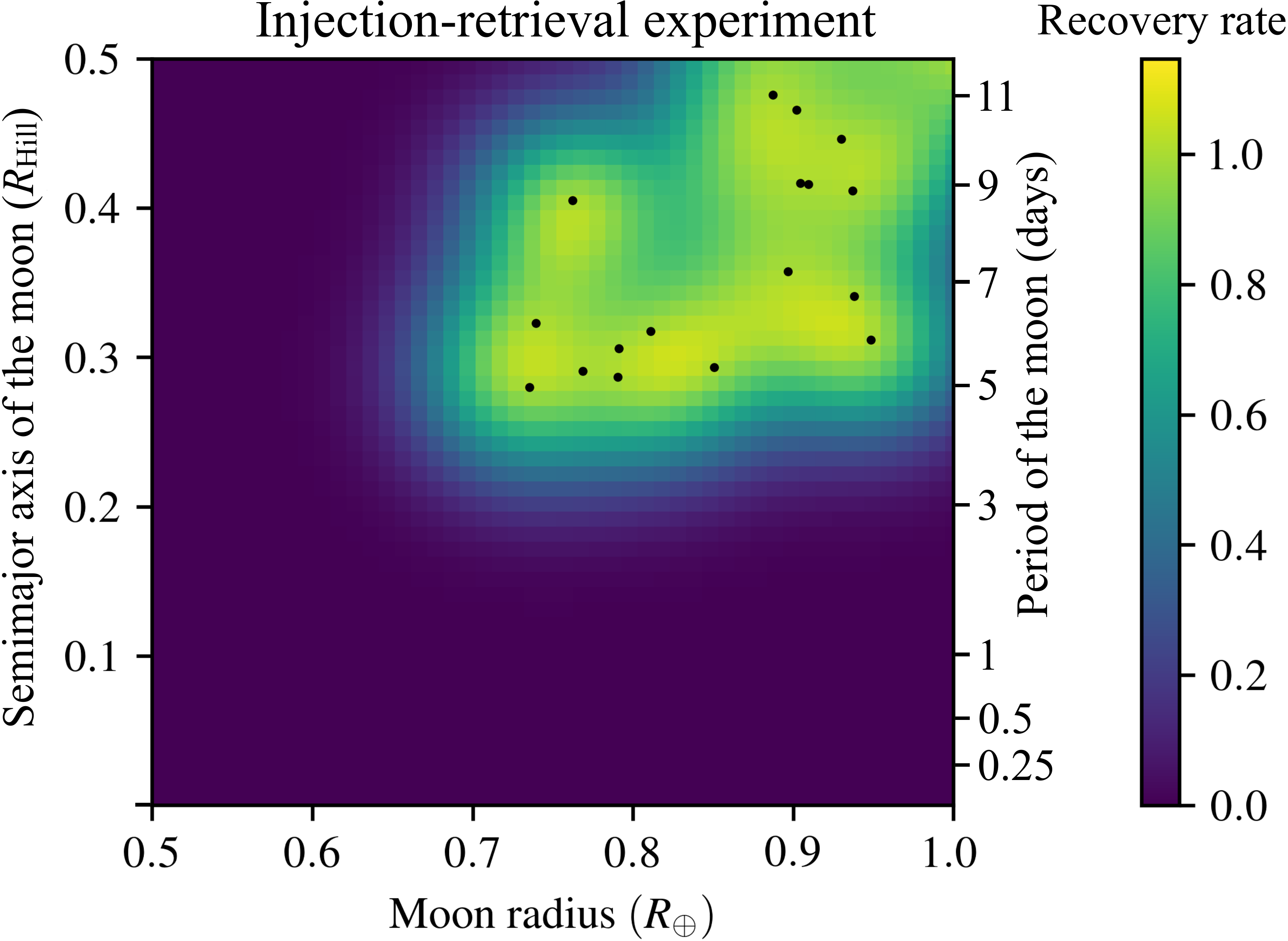}
\caption{Recovery rates of our simulated injection-retrieval experiment for different moon sizes and different orbits around a Neptune-sized planet. The planet and the moon are both in a 60\,d (0.3\,AU) orbit around a Sun-like star. Recovery is defined as successful (black dots) if $2\log_{\rm e}(B_{\rm mp})>9.21$. The color map represents the estimated density as a weighted sum of Gaussian distributions based on the successful recoveries. Recovery rates become substantial for planet-moon separation $>0.3\,R_{\rm Hill}$ (i.e. $P_{\rm m}>5.5\,$d) and moon radii $R_{\rm m}> 0.7 R_{\oplus}$.}
\label{fig:recovery}
\end{figure*}
%%%% Supplementary Fig. 6 %%%%

\clearpage

%%%% Supplementary Table 4 %%%%
\begin{table}[h]
\begin{center}
\caption{Evidence of an exomoon from our injection-retrieval experiments with Kepler data of Kepler-1625.}
\label{tab:injretr_K1625}%
\setlength{\tabcolsep}{3pt}
\begin{tabular}{l|c|c|c}
\toprule
$2 \log_e(B_{10})$  & Planet-only\footnotemark[1] & Planet-moon $2\sigma$-Pr.\footnotemark[2] & Planet-moon Copl.\footnotemark[3] \\
\midrule
\hspace{0.02cm} $<0$           & 96 (75.0\,\%)  &  1 (1.6\,\%)  &  9 (14.1\,\%)  \\
\hspace{0.273cm} $0$ -- $2.30$ &   6 (4.7\,\%)  &  0 (0.0\,\%)  &  1 (1.6\,\%)  \\
$2.30$ -- $4.61$               &   6 (4.7\,\%)  &  0 (0.0\,\%)  &  4 (6.3\,\%)  \\
$4.61$ -- $6.91$               &   2 (1.6\,\%)  &  1 (1.6\,\%)  &  1 (1.6\,\%)  \\
$6.91$ -- $9.21$               &   4 (3.1\,\%)  &  0 (0.0\,\%)  &  0 (0.0\,\%)  \\
\hspace{0.02cm} $>9.21$        &  14 (10.9\,\%) & 62 (96.9\,\%) & 49 (76.6\,\%) \\
\botrule
\end{tabular}
%\footnotetext{(any additional notes)}
\footnotetext[1]{Priors of the planet selected from the $2\sigma$ posteriors of our planet-only solutions. 128 systems were simulated.}
\footnotetext[2]{Priors of the injected exomoon drawn from the $2\sigma$ posteriors of our planet-moon model fits, except for the orbital phase, which was randomized, and the orbital period, which was drawn from a linear-spaced grid. 64 systems were simulated.}
\footnotetext[3]{Priors of the injected exomoon for coplanar orbits with randomized orbital phases and orbital periods selected from a linear-spaced grid. 64 systems were simulated.}
\end{center}
\end{table}
%%%% Supplementary Table 4 %%%%

\clearpage

%%%% Supplementary Table 5 %%%%
\begin{table}[h]
\begin{center}
\caption{Evidence of an exomoon from our injection-retrieval experiments with Kepler data of Kepler-1708.}
\label{tab:injretr_K1708}%
\setlength{\tabcolsep}{3pt}
\begin{tabular}{l|c|c|c}
\toprule
$2 \log_e(B_{10})$  & Planet-only\footnotemark[1] & Planet-moon $2\sigma$-Pr.\footnotemark[2] & Planet-moon Copl.\footnotemark[3] \\
\midrule
\hspace{0.02cm} $<0$           & 125 (97.7\,\%) & 22 (34.4\,\%) & 17 (26.6\,\%) \\
\hspace{0.273cm} $0$ -- $2.30$ &   0 (0.0\,\%)  &  3 (4.7\,\%)  &  3 (4.7\,\%)  \\
$2.30$ -- $4.61$               &   1 (0.8\,\%)  &  2 (3.1\,\%)  &  1 (1.6\,\%)  \\
$4.61$ -- $6.91$               &   0 (0.0\,\%)  &  2 (4.7\,\%)  &  3 (3.1\,\%)  \\
$6.91$ -- $9.21$               &   0 (0.0\,\%)  &  1 (1.6\,\%)  &  2 (3.1\,\%)  \\
\hspace{0.02cm} $>9.21$        &   2 (1.6\,\%)  & 34 (53.1\,\%) & 38 (59.4\,\%) \\
\botrule
\end{tabular}
%\footnotetext{(any additional notes)}
\footnotetext[1]{Priors of the planet selected from the $2\sigma$ posteriors of our planet-only solutions. 128 systems were simulated.}
\footnotetext[2]{Priors of the injected exomoon drawn from the $2\sigma$ posteriors of our planet-moon model fits, except for the orbital phase, which was randomized, and the orbital period, which was drawn from a linear-spaced grid. 64 systems were simulated.}
\footnotetext[3]{Priors of the injected exomoon for coplanar orbits with randomized orbital phases and orbital periods selected from a linear-spaced grid. 64 systems were simulated.}
\end{center}
\end{table}
%%%% Supplementary Table 5 %%%%

\clearpage

%%%% Supplementary Fig. 7 %%%%
\begin{figure}[h]%
\centering
\includegraphics[width=.995\textwidth]{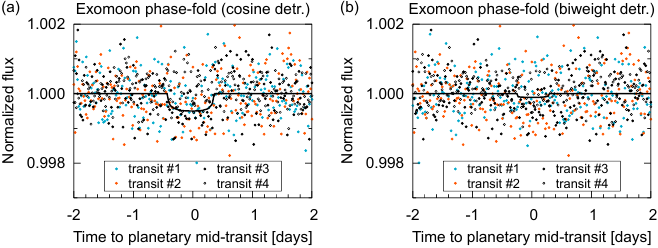}
\caption{Phase-folded light curve of the previously claimed moon transit of Kepler-1625\,b. The planetary transit component has been re-added to illustrate the residual moon transits from our best fit. Data of transits \#1-\#3 from Kepler and of transit \#4 from Hubble (points) as well as the best-fitting model for transit \#3 from Kepler (black line) were folded on the planetary mid-transit time. (a) Phased-folded moon transits for detrending approach 2 in which the stellar LDCs were fitted and light curve detrending was done using a sum of cosines filter. (b) Similar to (a) but using detrending approach 3 with a biweight filter. Details about the detrending approaches are given in the Methods.}
\label{fig:Kepler1625_fold}
\end{figure}
%%%% Supplementary Fig. 7 %%%%

\clearpage

%%%% Supplementary Fig. 8 %%%%
\begin{figure}[h]%
\centering
\includegraphics[width=.995\textwidth]{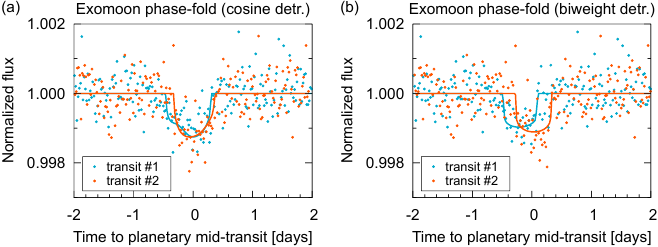}
\caption{Phase-folded light curve of the previously claimed moon transit of Kepler-1708\,b. The planetary transit component has been subtracted to illustrate the residual moon transits from our best fit. Data (points) and models (lines) from the first transit (blue) and second transit (orange) are folded on the planetary mid-transit time. The dip at $-1.5$\,d around the mid-transit time of transit \#2 is correlated noise and cannot be correlated with any possible exomoon. The amplitude of this noise is similar to that of the proposed exomoon transits and therefore possibly the cause of the statistical evidence. (a) LDCs were fitted as free parameters and light curve detrending was done using a sum of cosines filter. (b) Similar to (a) but using a biweight filter for detrending. Details about the detrending approaches are given in the Methods.}
\label{fig:Kepler1708_fold}
\end{figure}
%%%% Supplementary Fig. 8 %%%%

\clearpage

%%%% Supplementary Fig. 9 %%%%
\begin{figure*}[h]%
\centering
\includegraphics[width=.5\textwidth]{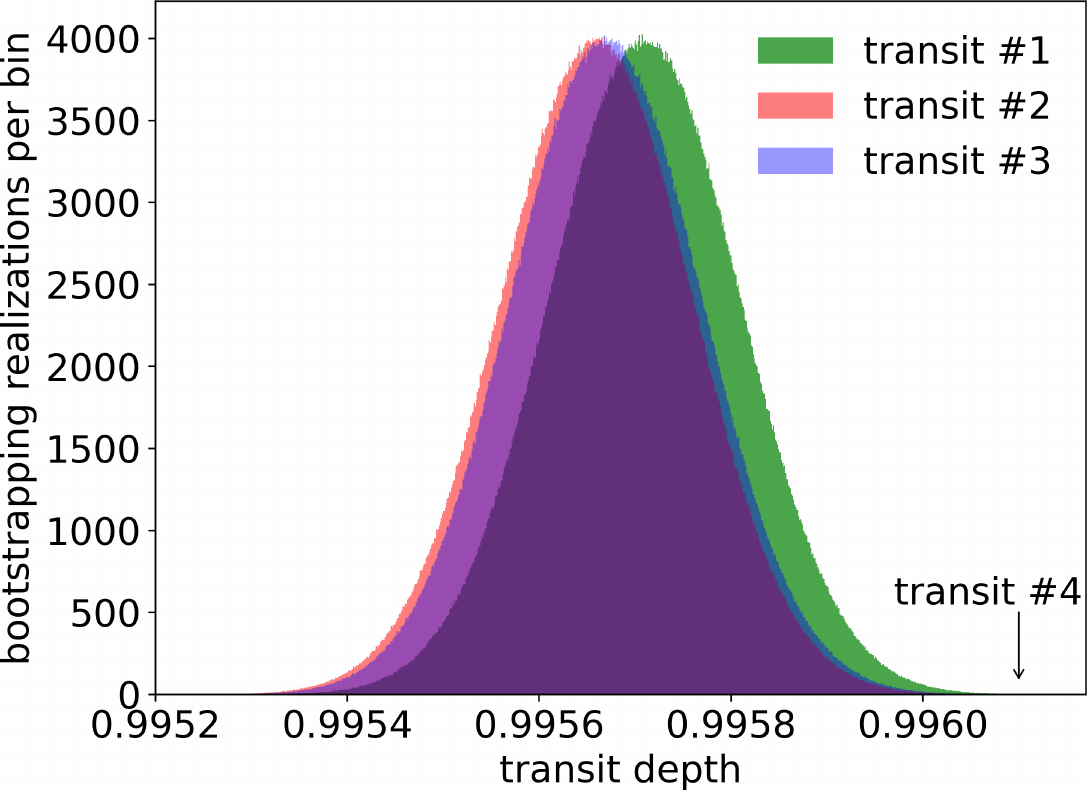}
\caption{Bootstrapping of ten million transit depth realizations for each of the three transits of Kepler-1625\,b from the Kepler mission. The depth of the transit observed with Hubble is indicated with an arrow as `transit \#4'.}
\label{fig:bootstrapping}
\end{figure*}
%%%% Supplementary Fig. 9 %%%%

\clearpage

%%%% Supplementary Fig. 10 %%%%
\begin{figure*}[h]%
\centering
\includegraphics[width=1\textwidth]{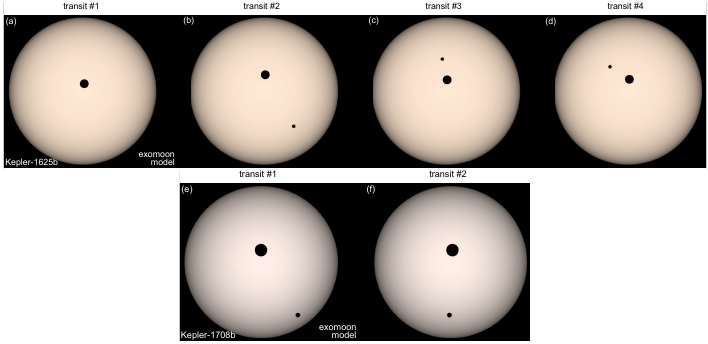}
\caption{Transit animations of Kepler-1625\,b (top) and Kepler-1708\,b (bottom) corresponding to the respective maximum likelihood planet-moon models. Each panel refers to the transit midpoint of the planet-moon barycenter. The full animations were generated with {\tt Pandora} and are available in the online Supplementary Materials.}
\label{fig:animation}
\end{figure*}
%%%% Supplementary Fig. 10 %%%%

\end{document}